\documentclass[%
 reprint,
 amsmath,amssymb,
 aps,
]{revtex4-2}

\usepackage{graphicx}% Include figure files
\usepackage{dcolumn}% Align table columns on decimal point
\usepackage{bm}% bold math
\renewcommand{\vec}[1]{\boldsymbol{\mathrm{#1}}}

\usepackage{xr}
\usepackage{color}

\begin{document}

\preprint{APS/123-QED}

\title{Long-Range Angular Correlations of Particle Displacements at a Plastic-to-Elastic Transition in Jammed Amorphous Solids}% Force line breaks with \\

\author{Yang Fu}
\affiliation{Beijing National Laboratory for Condensed Matter Physics and Laboratory of Soft Matter Physics, Institute of Physics, Chinese Academy of Sciences, Beijing 100190, China}
\author{Yuliang Jin}%
\email{yuliangjin@mail.itp.ac.cn}
\affiliation{%
Institute of Theoretical Physics, Chinese Academy of Sciences, Beijing 100190, China\\
School of Physical Sciences, University of Chinese Academy of Sciences, Beijing 100049, China\\
Center for Theoretical Interdisciplinary Sciences, Wenzhou Institute, University of Chinese Academy of Sciences, Wenzhou, Zhejiang 325001, China
}%

\author{Deng Pan}
\affiliation{
Institute of Theoretical Physics, Chinese Academy of Sciences, Beijing 100190, China
}%
\author{ Procaccia}
\affiliation{%
Sino-Europe Complex Science Center, School of Mathematics, North University of China, Shanxi, Taiyuan 030051, China\\
Department of Chemical Physics, The Weizmann Institute of Science, Rehovot 76100, Israel
}%

\date{\today}

\begin{abstract}
Understanding how a fluid turns into an amorphous solid is a fundamental challenge in statistical physics, 
during which no apparent structural ordering appears.
In the athermal limit, the two states are connected by a well-defined jamming transition, near which the solid is marginally stable. 
A recent mechanical response screening theory proposes an additional transition above jamming,  called a plastic-to-elastic transition here, separating  anomalous and quasielastic mechanical behavior. 
Through numerical inflation simulations in two dimensions, we show that the onsets of long-range radial and angular correlations of particle displacements decouple, occurring, respectively, at the jamming and plastic-to-elastic transitions. 
The latter is characterized by a power-law diverging correlation angle and a power-law spectrum of the displacements along a circle. 
This work establishes two-step transitions on the mechanical properties during ``decompression melting'' of an athermal overjammed amorphous solid, reminiscent of the two-step structural melting of a crystal in  two dimensions. In contradistinction with the latter, the plastic-to-elastic transition exists also in three dimensions.
\end{abstract}

\maketitle

{\bf Introduction.}
When strained by an increasing deformation $\gamma$, a crystal displays a crossover from  elastic to plastic responses. If the crystal is compressed, it responds  elastically to a small $\gamma$ at any pressure $p$. The picture is dramatically changed in athermal amorphous solids near the jamming transition ($p \approx 0$). Such a solid is known to be marginally stable~\cite{liu2010jamming, muller2015marginal, charbonneau2014fractal}, with elasticity breaking down in the thermodynamic limit, even to infinitesimal mechanical perturbations~\cite{hentschel2011athermal, biroli2016breakdown}. 
Indeed, the minimum strain $\gamma_{\rm min}$ required to trigger a plastic event vanishes in large systems~\cite{karmakar2010statistical, morse2020differences}.
On the other hand, one expects  elasticity restored in overjammed solids well above the jamming transition ($p \gg 0$), when the interparticle overlapping $\delta$ exceeds expected $\gamma_{\rm min}$, such that under small deformations the system can be effectively considered as an elastic medium.
%contact network remains unchanged. 
Thus there should be a {\it plastic-to-elastic (PE) transition} (or crossover) moving away from jamming. 
 
The expected PE transition has been recently explored by two approaches, both of which originated from theories. The first approach builds on the mean-field replica theory (MFRT)~\cite{charbonneau2014fractal, parisi2020theory}, which predicts a Gardner transition~\cite{charbonneau2014fractal, charbonneau2014exact, charbonneau2015numerical,  berthier2016growing, urbani2023gardner} 
in athermal soft spheres at $\varphi_{\rm G}^>$ above the jamming density $\varphi_{\rm J}$~\cite{biroli2016breakdown}. The Gardner transition separates the elastic phase and the plastic phase where the shear modulus is protocol dependent due to marginal stability~\cite{yoshino2014shear, jin2017exploring, jin2018stability, nakayama2016protocol}.

\begin{figure}[thbp]
%\begin{figure*}[thbp]
  \centering
\includegraphics[width=1\linewidth]{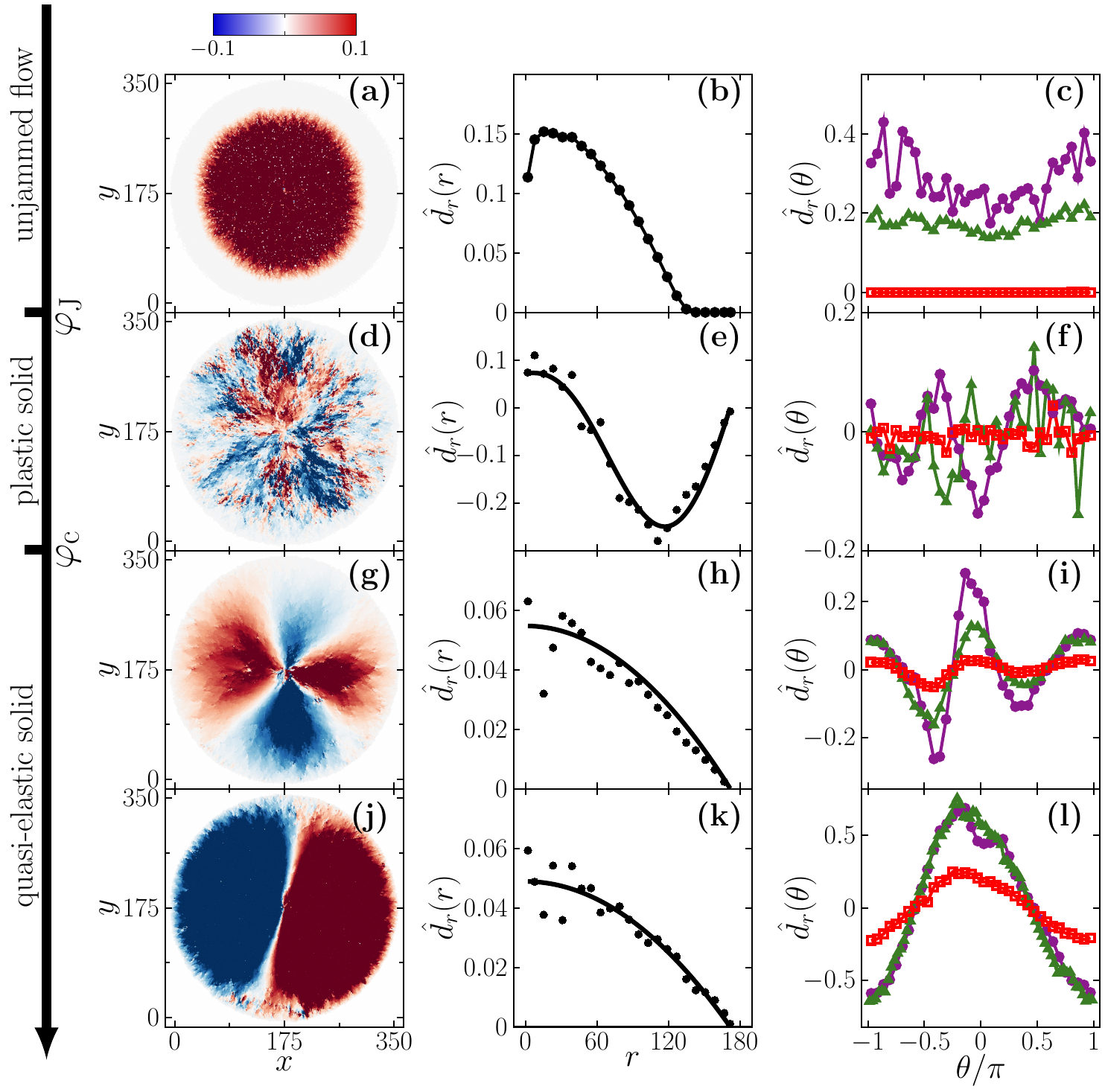}
    \caption{ {\bf Radial displacement fields.}
    Each row corresponds to typical response behavior at different pressures  (from top to bottom): $p=0, 4.4 \times 10^{-2}, 5.2 \times 10^2, 8.6\times 10^2$, for $N=67352$. 
    First column: heat maps of  $\hat{d}_r(x,y)$ normalized by the maximum $\hat{d}_r^{\rm max}$.
    Second column: $\hat{d}_r(r)$, where
    the lines in (h,k) represent Eq.~(\ref{Eq:dr_elastic}) with  $d_0^{\rm eff}=0.054$ and 0.048, respectively, and the line in (e) represents Eq.~(\ref{Eq:dr_anomolous}) with $d_0^{\rm eff} = 0.072$ and $\kappa_{\rm fit} = 0.026$. 
    Third column: $\hat{d}_r(\theta)$ at $\hat{r} \equiv r/r_{\rm out} = 0.15, 0.5$, and $0.85$
    (purple, green, and red, respectively). 
    }
	\label{Fig:displacement}
%\end{figure*}
\end{figure}

The second approach is motivated by a recent mechanical response screening theory (MRST)~\cite{lemaitre2021anomalous, mondal2022experimental,  bhowmik2022direct,  kumar2022anomalous, charan2023anomalous, kumar2024elasticity}. The theory provides a possible PE transition in athermal overjammed solids, between a quasielastic phase with quadrupole screening of the elastic field by plastic events, where plasticity simply renormalizes the elastic moduli, and an anomalous phase with dipole screening. This transition is observed in recent simulations, by showing the change of the screening parameter $\kappa$ from zero to a finite value at a nonzero pressure $p_{\rm c}$~\cite{jin2023intermediate}.
However, several crucial questions remain to be answered regarding the nature of the transition: in particular, whether a sharp transition or a crossover occurs at $p_{\rm c}$ in the thermodynamic limit, and in the former case, whether it is associated with diverging long-range correlations.

Following the second approach, here we show that the PE transition can be unambiguously differentiated from the jamming transition:
%, although a finite-size scaling $p_{\rm c} \sim N^{-3/2}$ (for Hertzian interactions) suggests that $p_{\rm c}$ coincides with the jamming transition $p_{\rm J}=0$ in the thermodynamic limit. Physically, 
the jamming transition is associated with a diverging length due to long-range radial correlations in displacements, and, in contrast, the PE transition corresponds to a diverging angle due to long-range transverse correlations. It should be noted that the PE transition captures the change of the  macroscopic mechanical response behavior, which is beyond the first contact change events at the microscopic level~\cite{van2014contact, van2016contact}.

{\bf Three phases: quasielastic, plastic and unjammed.}
We simulate a bidisperse granular model of Hertzian (unless otherwise specified) particles in two dimensions, confined between a fixed outer boundary of radius $r_{\rm out}$ and an inner boundary of radius $r_{\rm in}$. A small particle, with a unit diameter ($R_1 =0.5$) and a fixed position at the center ($r=0$), is inflated as $r_{\rm in} \to r_{\rm in} + d_0$, where $r_{\rm in} = R_1 = 0.5$ and $d_0 = 0.1$ (unless otherwise specified).
The energy of the system is minimized after this instantaneous inflation, and the particle responses are characterized by the displacement field $ \vec{d}(r, \theta) $.
In this study, we focus on the radial component of the displacement $ d_r(r, \theta) $, where $d_r = \vec{d} \cdot \hat{\vec{n}}$ and $\hat{\vec{n}}$ is the unit radial vector.
%~\YF{$\hat{r}$ is defined again later as $\hat{r}=r/r_{\rm{out}}$}.
The transverse component $ d_{\theta} (r, \theta) $ is investigated in a separate study~\cite{fu2024odd}, revealing an interesting {\it odd dipole screening} effect.
Details of the model and simulation methods are described in Appendixes A and B.

The mechanical responses to the above-described inflation can be categorized into three phases:

(i) A quasielastic solid phase at $p> p_{\rm c}$ and $\varphi> \varphi_{\rm c}$, where the mechanical response in the radial direction is elasticlike. The displacement field is dominated by dipolar and quadrupolar patterns [see Figs.~\ref{Fig:displacement}(g)-1(l)].
Although the central inflation is isotropic, the actual displacement and force fields are generally anisotropic.
%, revealing an important difference between a continuous medium and a discrete particle packing.In other words, the dipolar and quadrupolar patterns reflect disorder-induced breaking of rotational symmetry in granular models. 
Hexapoles and higher-order poles exist but are originated from $1/f^{1.5}$ noise (see Fig.~\ref{Fig:Correlation_PS} and related discussions).

The displacement field $d_r(r, \theta)$ in this phase is consistent with the Michell solution that describes the mechanical response of a standard elastic medium~\cite{Kumar_2023}.
Using the rescaled radial displacements $\hat{d}_r (r, \theta) = r d_r (r, \theta)$ to remove the trivial $1/r$ decay  of $d_r (r, \theta)$ at large $r$ and averaging out the angular dependence, the radial dependence of $\hat{d}_r(r)$ is expected to follow the solution of the normal elasticity theory,
\begin{align}
\hat{d}_r(r) = d_0 \frac{r_{\rm in}(r^2-r_{\rm out}^2)}{(r_{\rm in}^2-r_{\rm out}^2)}.
\label{Eq:dr_elastic}
\end{align}
Eq.~(\ref{Eq:dr_elastic}) is consistent with simulation results, with the inflation variable treated as a fitting parameter $d_0 \to d_0^{\rm eff}$ [see Figs.~\ref{Fig:displacement}(h) and 1(k)].

Eq.~(\ref{Eq:dr_elastic}) suggests that the response is elasticlike. The application of inflation may result in plastic events, which are typically quadrupolar in nature, aka {\it Eshelby inclusions}. At a large $p$, the plastic events are sparse, and the gradient of the quadrupole field $\vec{Q}(r, \theta)$ is negligible. In this case, the effect of plasticity is to renormalize the elastic moduli, without modifying the overall behavior of the displacement field Eq.~(\ref{Eq:dr_elastic}).
For this reason, the high-pressure phase is termed a {\it quasielastic phase}. However, the angular dependence of $\hat{d}_r$ is not isotropiclike in a true elastic medium (see Sec.~S1 in Supplemental Material (SM)~\cite{fu2025SI}). 
%~\YF{seen in \ref{SI: Dipolar and quadrupolar} \cite{fu2025SI}}. 
The oscillative behavior of $\hat{d}_r(\theta)$ (at a given $r$) suggests long-range correlations in the angular direction [Figs.~\ref{Fig:displacement}(i) and 1(l)).

 (ii) A plastic solid phase 
at $0<p<p_{\rm c}$ and $\varphi_{\rm J} < \varphi < \varphi_{\rm c}$, where
the system is overall jammed but the mechanical response anomalously disobeys elasticity. 
The displacement field $\hat{d}_r(r, \theta)$ can not be described by any regular patterns [see Figs.~\ref{Fig:displacement}(d)-(f)]. 
Under the isotropic assumption, 
the screening theory provides a solution of $\hat{d}_r(r)$~\cite{lemaitre2021anomalous}, 
\begin{align}
\hat{d}_r(r) = d_0 r \frac{Y_1(r \kappa) J_1(r_{\rm out} \kappa) - J_1(r \kappa) Y_1(r_{\rm out} \kappa)}{Y_1(r_{\rm in} \kappa) J_1(r_{\rm out} \kappa) - J_1(r_{\rm in} \kappa) Y_1(r_{\rm out} \kappa)},
\label{Eq:dr_anomolous}
\end{align}
where $J_1$ and $Y_1$ are circular Bessel functions of the first and second kind, respectively, and $\kappa$ is a {\it screening parameter}. Simulation data are fitted to Eq.~(\ref{Eq:dr_anomolous}) with two fitting parameters, $d_0^{\rm eff}$ and  $\kappa_{\rm fit}$ [see Fig.~\ref{Fig:displacement}(e)]. Note that Eq.~(\ref{Eq:dr_elastic}) is equivalent to setting $\kappa = 0$ in Eq.~(\ref{Eq:dr_anomolous}). See Sec.~S2 in SM~\cite{fu2025SI, ellenbroek2006critical, ellenbroek2009jammed} for ensemble-averaged data.
%~\YF{seen in \ref{SI: Ensemble-averaged radial displacements} \cite{fu2025SI}}

The screening parameter $\kappa \sim 1/\ell_{\rm s}$ is the inverse of the {\it screening length} $\ell_{\rm s}$. The screening effect originates from  an effective dipole field $\vec{P}(r, \theta)$, which is the gradient of $\vec{Q}(r, \theta)$ when the quadrupolar events are dense and non-uniformly distributed in the space~\cite{bhowmik2022direct}. 
While Eq.~(\ref{Eq:dr_anomolous}) gives rise to long-range correlations in the radial direction, $d_r(\theta)$ is noiselike, implying short-range correlations in the angular direction [Fig.~\ref{Fig:displacement}(f)].

(iii) An unjammed phase at $p=0$ and $\varphi < \varphi_{\rm J}$, where the system is in a  
fluid state. The influence of the central inflation propagates only up to a finite distance $\xi$, and $d_r(\theta)$ is noiselike [see Figs.~\ref{Fig:displacement}(a)-1(c)].

\begin{figure}[htbp]
  \centering
\includegraphics[width=1\linewidth]{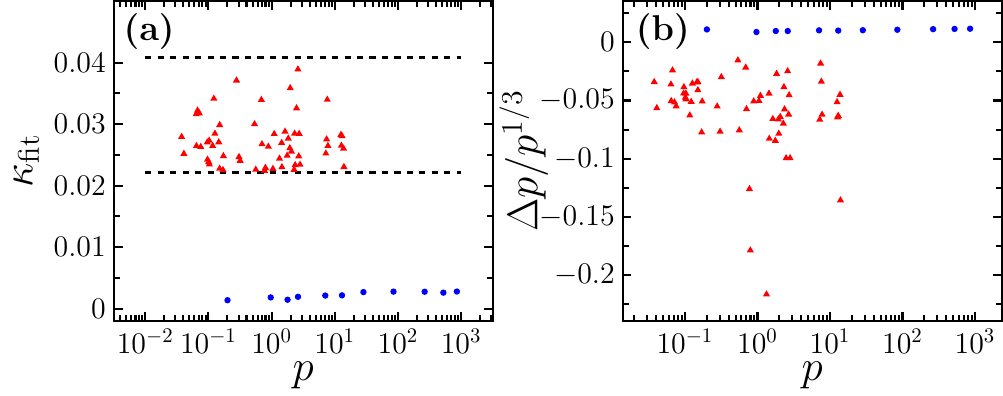}
    \caption{{\bf Plastic-to-elastic transition.}
    (a) The screening parameter $\kappa_{\rm fit}$ from fitting, and (b) the scaled pressure difference $\Delta p/p^{1/3}$ as  functions of $p$, for $N=67352$.
    The blue data points (averaged over samples with $\kappa_{\rm fit}<0.01$ at the given $p$; error bars are invisible) have near-zero values, which can be unambiguously distinguished from others (red points; each point represents one sample).
    }
	\label{Fig:2PT}
\end{figure}

Let us highlight the essential differences between the three phases by summarizing the radial and angular dependence of $\hat{d}_r$ (see Fig.~\ref{Fig:displacement}):
(i) in the quasielastic solid phase, $\hat{d}_r(r)$ follows the elasticity theory Eq.~(\ref{Eq:dr_elastic}) and $\hat{d}_r(\theta)$ is oscillative (both long ranged); (ii) in the plastic solid phase, $\hat{d}_r(r)$ follows the MRST Eq.~(\ref{Eq:dr_anomolous}) and $\hat{d}_r(\theta)$ is noiselike (long ranged in $r$ and short ranged in $\theta$); (iii) in the unjammed phase, $\hat{d}_r(r)$ decays to zero at a finite $\xi$ and $\hat{d}_r(\theta)$ is noiselike (both short ranged). 
%Below we characterize these differences in detail, focusing on the features of the two transitions between the three phases. 

 {\bf Two transitions: jamming and plastic-to-elastic.}
The jamming transition between unjammed and plastic phases  has been extensively investigated in previous studies~\cite{makse2000packing, o2003jamming,   olsson2007critical, heussinger2009jamming, van2009jamming,  behringer2018physics, liu2010jamming, pan2023review}. In Ref.~\cite{heussinger2009jamming}, the radial correlations between particles' nonaffine displacements are measured in athermal quasistatic shear simulations with Lees-Edwards boundary conditions, giving a diverging correlation length $\xi$ when the jamming transition is approached from below, 
\begin{align}
\xi \sim (\varphi_{\rm J} - \varphi)^{-\nu},
\label{eq:xi}
\end{align}
with $\nu \sim 0.8 - 1.0$. 
In this study, we estimate $\xi$ by the length scale at which the radial displacement $d_r(r)$ decays to a small threshold value, $d_r(r=\xi) = 10^{-3}$ (see Sec.~S3 in SM \cite{fu2025SI}).
%(see Appendix Fig.~\ref{Fig:jamming}\textit{a}).
%\YF{In this study, we estimate $\xi$ by the length scale at which the radial displacement almost vanishes.
%~\YF{(see Fig.~\ref{Fig:jamming}\textit{a} \cite{fu2025SI}}
The power-law divergence of $\xi$ is confirmed
%in Appendix  Fig.~\ref{Fig:jamming}\textit{b} ~\YF{\cite{fu2025SI}}, 
with $\varphi_{\rm J} = 0.842$~\cite{o2002random} and $\nu \approx 0.8$~\cite{heussinger2009jamming} consistent with previous studies.

The PE transition between plastic and  quasielastic phases was only noticed very recently~\cite{jin2023intermediate}. To provide a rough estimation of the PE transition pressure $p_{\rm c}$, we follow the strategy in ~\cite{jin2023intermediate}: the data of $\hat{d}_r(r)$ at different $p$ are fitted to Eq.~(\ref{Eq:dr_anomolous}), and the  fitting  parameter $\kappa_{\rm fit}$  is plotted as a function of $p$ in Fig.~\ref{Fig:2PT}(a).
At high $p$, $\kappa_{\rm fit} \approx 0$ (blue points), consistent with Eq.~(\ref{Eq:dr_elastic}). The nonzero $\kappa_{\rm fit}$ appears only below $p_{\rm c} \approx 20$ for the system of $N=67352$ particles (red points).
In Ref.~\cite{jin2023intermediate}, a {\it selection principle} of $\kappa$ is proposed: 
the $\hat{d}_r(r)$ curves measured in simulations preferably select a series of $\kappa_n  \approx n \pi/r_{\rm out}$  with a decreasing probability when $n$ is increased, where $n=1,2,\ldots$
At $\kappa = \kappa_n$ the screening effects are significantly strong, which can be seen from the behavior of Eq.~(\ref{Eq:dr_anomolous}), since the  amplitude $\hat{d}_{\rm amp} (\kappa = \kappa_n)$ diverges. Here $\hat{d}_{\rm amp} (\kappa)$ is the extreme value of  the $\hat{d}_r(r)$ curve for the given $\kappa$.
In Fig.~\ref{Fig:2PT}(a), the  first two $\kappa_n$, i.e., $\kappa_1 \approx 0.022$ and $\kappa_{2} \approx 0.041$, are indicated by dashed lines.  The data points of nonzero $\kappa_{\rm fit}$ all fall in the range between $\kappa_1$ and $\kappa_2$. The clear gap between $\kappa = 0$ and $\kappa_1$ supports the selection principle. Meanwhile, we observe a wide fluctuation of $\kappa_{\rm fit}$ at $p<p_{\rm c}$, possibly due to strong sample-to-sample fluctuations in finite-sized systems and the uncertainty in the fitting.

\begin{figure}[thbp]
  \centering
\includegraphics[width=1\linewidth]{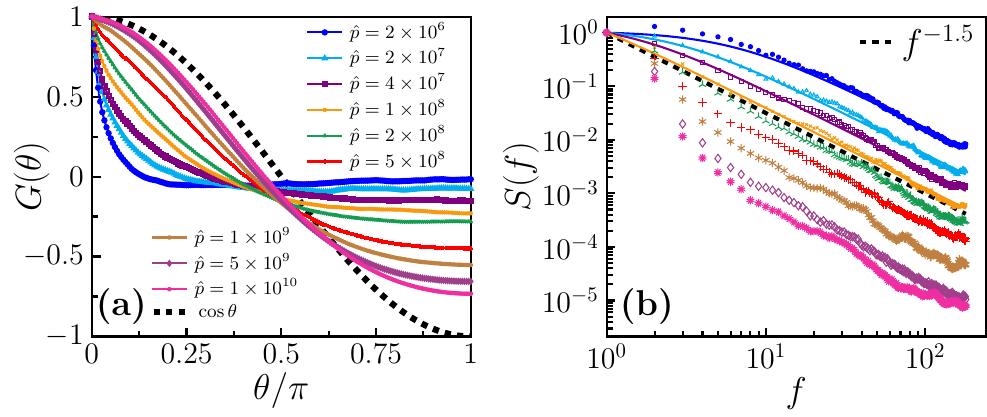}
    \caption{{\bf Angular correlation functions and power spectrum}.
    (a) Angular correlation function $G(\theta)$ and (b) power spectrum $S(f)$ at different $\hat{p}= p N^{3/2}$, for $\hat{r} =r/r_{\rm out} = 0.5$ and $N=67352$.
    The solid lines in (b) represent fitting to
    Eq.~(\ref{eq:Lorentzian}).}
	\label{Fig:Correlation_PS}
\end{figure}

To examine the reliability of $p_{\rm c}$ estimated above, in this study we propose two additional methods to determine $p_{\rm c}$. In Fig.~\ref{Fig:2PT}(b), we plot the pressure difference $\Delta p \equiv p_{\rm final}-p$ between the pressures after and before the central inflation (both pressures are measured with the energy minimized). 
To properly normalize the pressure, note that $p \sim F \sim \delta^{3/2}$, where the second relation between the interparticle force $F$ and the interparticle linear overlap $\delta$ holds for the Hertzian potential. From this, one obtains  $d \delta \sim dp/p^{1/3}$. 
The data (red points) in Fig.~\ref{Fig:2PT}(b) shows that $\Delta p/p^{1/3}$ fluctuates around $-0.05$  below  $p_{\rm c} \approx 20$.
The fluctuation is large, but nevertheless the negative values confirm the anomaly, compared to the 
small positive values expected from the elastic behavior (blue points). Note that this second approach can be straightforwardly accessed in granular experiments~\cite{mondal2022experimental}.  
In the next section, we will discuss a third way to determine $p_{\rm c}$, which gives the most accurate estimate.

%The above analysis shows that jamming and PE transitions are two separate transitions, at least in finite-sized systems. For a finite $N$, the PE transition occurs at a finite $p_{\rm c}$, while the jamming transition occurs at $p=0$ for any $N$. A finite-size analysis will be presented below to extrapolate the asymptotic behavior in the thermodynamic limit. 

\begin{figure}[thbp]
  \centering
\includegraphics[width=1\linewidth]{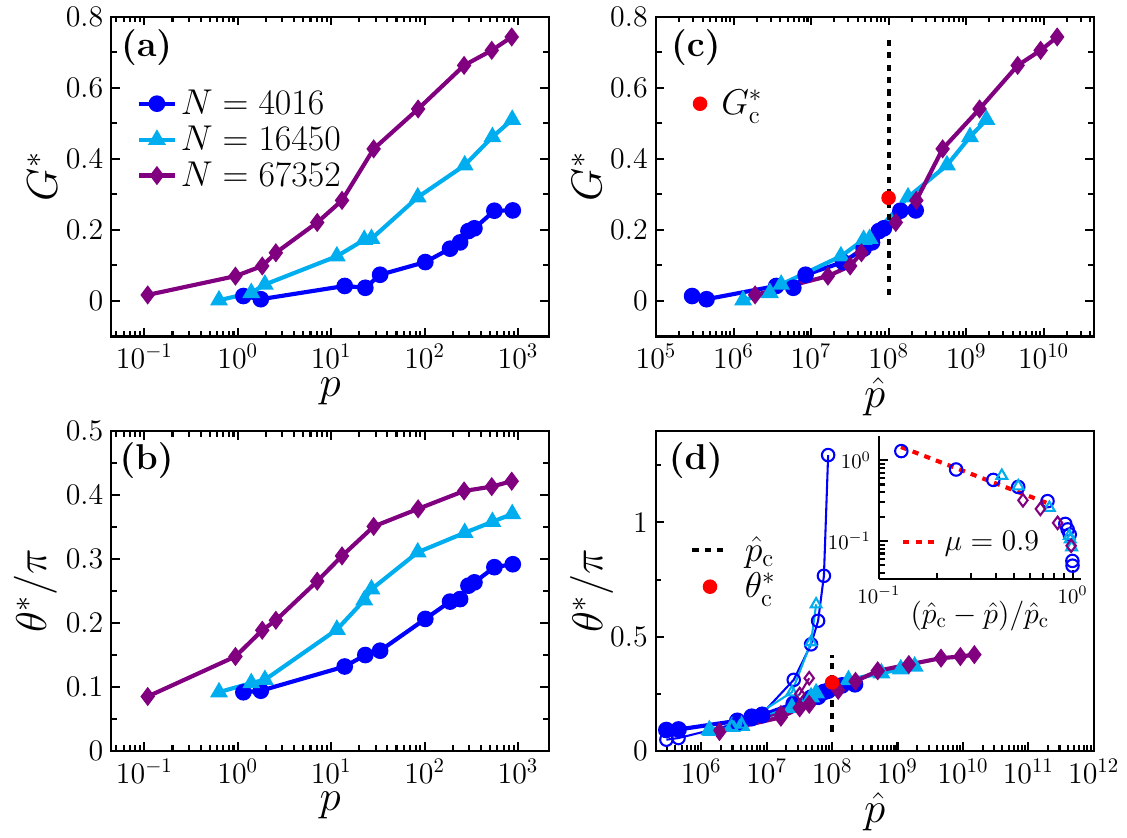}%{order_parameter_correlation_length.pdf}
    \caption{{\bf Order parameters and correlation angles.}
    $G^*$ and $\theta^*$  as functions of (a,b) $p$ and (c,d) $\hat{p}$, for $\hat{r} = 0.5$ and different $N$.
    Open symbols in (d) are
   $\theta^\dagger$ data.
    (inset) $\theta^\dagger$ as a function of $(\hat{p}_{\rm c} - \hat{p})/\hat{p}_{\rm c}$, with the power-law fitting to Eq.~(\ref{eq:theta_dagger}).
     }
	\label{Fig:Order_parameter}
\end{figure}

{\bf Emergence of long-range angular correlations at the plastic-to-elastic transition.}
As discussed above, the onset of long-range displacement correlations in the radial direction occurs at the jamming transition $p = p_{\rm J}=0$ and $\varphi = \varphi_{\rm  J}$.
%~\red{\cite{fu2025SI}} ~\YF{seen in \ref{SI: Jamming transition} \cite{fu2025SI}}.
%(Appendix Fig.~\ref{Fig:jamming} ~\YF{\cite{fu2025SI}}). 
Next we show that the onset of long-range displacement correlations in the angular direction appears at the PE transition $p= p_{\rm c}$. 
We define an angular correlation function of the radial displacements,
\begin{align}
G(\theta) = \overline{c\langle \delta{d}_r(\theta_0) \, \delta{d}_r(\theta_0+\theta)\rangle}, 
\label{eq:Angular_correlation_function}
\end{align}
where 
$\delta {d}_r(\theta) = {d}_r(\theta) - \langle d_r \rangle$ with $\langle d_r \rangle = \frac{1}{2 \pi} \int_0^{2\pi} {d}_r(\theta_0)  d\theta_0$, $c =1/ \langle |\delta{d}_r(\theta_0) |^2 \rangle$ is the normalization constant, and $\overline{x}$ represents the average over samples.
The correlation function $G(\theta)$ is computed from  $\delta{d}_r$ of particles within an annulus located at $[\hat{r}-\delta \hat{r}, \hat{r}+\delta \hat{r}]$, where $\hat{r} = r/r_{\rm out}$ (recall that $r_{\rm in} \ll r_{\rm out}$), and $\delta \hat{r} = 0.03$ is a small parameter.

Figure ~\ref{Fig:Correlation_PS}(a) reports $G(\theta)$ for
$\hat{r} = 0.5$ at different pressures.
In the large pressure limit, the  displacement field is dominated by dipolar patterns as shown in Fig.~\ref{Fig:displacement}(j), which corresponds to $G(\theta) = \cos(\theta)$.
With decreasing $p$, quadrupolar and other patterns [see Fig.~\ref{Fig:displacement}(g)]  appear, and  $G(\theta)$ deviates from the cosine function. Below $p_{\rm c}$, $G(\theta)$ rapidly decays to zero, indicating short-range correlations. To quantify the behavior change of $G(\theta)$,  we define an order parameter $G^* \equiv - G(\theta = \pi)$, and a correlation angle  $G(\theta =\theta^*) = G_{\rm th}$ with a threshold $G_{\rm th}  = 0.1$~(see Sec.~S4 in SM \cite{fu2025SI}). Figures ~\ref{Fig:Order_parameter}(a) and \ref{Fig:Order_parameter}(b) show that both $G^*$ and $\theta^*$ grow with $p$.

The data of $G^*$ and $\theta^*$ suffer strong finite-size effects and thus it is necessary  to consider a proper finite-size scaling. 
We find that they can be collapsed when plotted as functions of a rescaled pressure,  $\hat{p} = p N^{3/2}$ [Figs.~\ref{Fig:Order_parameter}(c) and ~\ref{Fig:Order_parameter}(d)]. To understand this finite-size scaling, recall that near the jamming transition, the isostatic length $\ell_{\rm iso} \sim (Z-Z_{\rm iso})^{-1} \sim p^{-1/3}$ sets a natural length scale, where $Z$ is the average coordination number and $Z_{\rm iso} = 2d = 4$ is the isostatic coordination number in $d=2$ dimensions~\cite{wyart2005geometric}. 
According to the argument in Ref.~\cite{jin2023intermediate}, the PE transition occurs when $\ell_{\rm iso}$ matches the largest screening length $\ell_{\rm s} \sim 1/\kappa_1 \approx r_{\rm out}/\pi \sim N^{1/2}$, which gives $p_{\rm c} \sim N^{-3/2}$, consistent with our data (see SM ~\cite{fu2025SI}).
For a more general interaction potential $U(\delta) \sim \delta^\alpha$, $p_{\rm c}(N,d_0)$ of different $N$ and inflation $d_0$ obeys a scaling function $N^\beta p_{\rm c}(N,d_0) = \mathcal{G}(N^2 d_0)$, where $\beta = 2(\alpha-1)$ and $\mathcal{G}(x) \sim x^{\beta/4}$~(see Appendix D and Sec.~S5 in SM \cite{fu2025SI}).
%Note that the finite-size scaling $p_{\rm c} \sim N^{-3/2}$ holds when $r_{\rm in}$ is fixed and $r_{\rm out} \sim N^{1/2}$ is varied; if one fixes the ratio $r_{\rm out}/r_{\rm in}$, then $p_{\rm c}$ is nearly independent of $N$~\cite{jin2023intermediate}.

%Note that this scaling holds only for  Hertzian systems. For Hookean systems, it becomes $p_{\rm c} \sim N^{-1}$ (Appendix  Fig.~\ref{fig:Hookean} ~\YF{\cite{fu2025SI}}). {In Appendix E ~\YF{In the Secs. S2 and S3 \cite{fu2025SI}}, we additionally examine effects of the inflation $d_0$, and find that $p_{\rm c} \sim d_0^{3/4}$. ~\YF{Rewrite?}}

Next we demonstrate that the system has long-range angular correlations at $\hat{p}_{\rm c}$, even though $\theta^*$ does not appear to diverge. In our setup, the long-range correlation corresponds to the convergence, $\theta^* \to \theta^*_{\rm c} \approx 0.3 \pi$ and  $G^* \to G^*_{\rm c} \approx 0.3$,
when $\hat{p} \to \hat{p}_{\rm c}$
from below.  
To show that, we perform the Fourier transform of $d_r(\theta)$ and plot the (normalized) power spectrum $S(f)$ in Fig.~\ref{Fig:Correlation_PS}(b) at different $\hat{p}$, for $N=67352$. Interestingly, $S(f)$ displays a power-law with an exponent $-1.5$ (which we term {\it $1/f^{1.5}$ noise}) at $\hat{p}_{\rm c} \approx 10^8$, indicating angular scale-free behavior.
%that the system is scale-free at the transition.
The determined $\hat{p}_{\rm c} \approx 10^8$ is consistent with $\hat{p}_{\rm c} \approx 3 \times 10^8$ estimated in Fig.~\ref{Fig:2PT} (considering strong data fluctuations there) and  $\hat{p}_{\rm c} \approx 8 \times 10^7$ reported in Ref.~\cite{jin2023intermediate} (converted from $p_{\rm c}\approx 3.5$ for $N=80000$). 
In the analyses below we take $\hat{p}_{\rm c} \approx 10^8$ as the value of the rescaled transition pressure. 
In Sec.~S6 of SM~\cite{fu2025SI}, 
%~\YF{\ref{SI: Dependence of the power spectrum on the radius} \cite{fu2025SI}}, 
we show that the $1/f^{1.5}$ behavior at $\hat{p}_{\rm c}$ is robust against varying $\hat{r}$, and is universal in Hertzian and Hookean systems.

With the power-law behavior at $\hat{p}_{\rm c}$ established, it becomes possible to characterize the growth of angular correlations from the spectrum data. We fit the $S(f)$ data at $\hat{p}< \hat{p}_{\rm c}$ to a generalized Lorentzian form,
\begin{align}
S_{\rm GL}(f) = \frac{A}{1+(f/f^{\dagger})^{1.5}},
\label{eq:Lorentzian}
\end{align}
where $f^{\dagger}$ is the only fitting parameter, and $A = 1+(1/f^{\dagger})^{1.5}$ is fixed by $f^{\dagger}$ due to the normalization $S(f=1) = 1$. 
Here $f^{\dagger}$ separates the white noise at $f<f^{\dagger}$ with $S(f) \sim 1$ from the $1/f^{1.5}$ noise at $f>f^{\dagger}$.
Since $f^{\dagger}$ is a characteristic wave number, one can define a corresponding angle, $\theta^\dagger \equiv k/f^\dagger$, with $k = 0.45$ estimated in Sec.~S7 of SM~\cite{fu2025SI, kubo2012statistical}.
%~\YF{\ref{SI: Inverse Fourier transform} \cite{fu2025SI}}.
%$\theta^{\dagger} \sim 1/f^{\dagger}$.
%\red{In practice, we define $\theta^\dagger \equiv k/f^\dagger$ with $k=81$ (see SM Sec.~S6 for details).}
The physical meaning of $\theta^{\dagger}$ can be interpreted as follows: the angular correlation between displacements exists up to $\theta^{\dagger}$, beyond which the displacements are uncorrelated like 
white noise.
%Thus $\theta^{\dagger}$ is another measure of the correlation angle, and we expect $\theta^{\dagger} \sim \theta^* \sim 1/f^{\dagger}$ for large $f^{\dagger}$.
%Since one can simply 
Assuming uncorrelated displacements in the fluid state, one expects that $\theta^{\dagger} \to 0$
%$f^{\dagger} \to \infty$ (or $\theta^{\dagger} \to 0$) 
approaching the jamming transition.
%, $\hat{p} \to \hat{p}_{\rm J}=0$.
On the other hand, %an elastic medium 
the scale-free spectrum $S(f)$ at $\hat{p}_{\rm c}$ suggests 
%should have long-range correlated displacements, which means that 
that $\theta^\dagger \to \infty$ (i.e., $f^{\dagger} \to 0$) approaching the PE transition.
Moreover, in the plastic  phase, 
%$\theta^\dagger = k/f^{\dagger}$ is finite, and 
$S(f)$ deviates from the scale-free scaling due to a finite $\theta^\dagger$. 
These expectations are consistent with the data in Fig.~\ref{Fig:Correlation_PS}(b). In particular, near $\hat{p}_{\rm c}$, $\theta^{\dagger}$ diverges such that
\begin{align}
\theta^{\dagger} \sim (\hat{p}_{\rm c} - \hat{p})^{-\mu},
\label{eq:theta_dagger}
\end{align}
where $\mu \approx 0.9$ from fitting [see Fig.~\ref{Fig:Order_parameter}(d) (inset)]. Eq.~(\ref{eq:theta_dagger}) is the angular counterpart of Eq.~(\ref{eq:xi}). Note that $\theta^{\dagger}$ diverges at $p_{\rm c}$ (or $\varphi_{\rm c}$) and $\xi$ diverges at $p_{\rm J}$ (or $\varphi_{\rm J}$).

The two correlation angles, $\theta^*$ and $\theta^{\dagger}$, coincide at small $\hat{p}$ [Fig.~\ref{Fig:Order_parameter}(d)].
The angle  $\theta^*$ does not diverge at $p_{\rm c}$, because it is defined in a finite domain $\theta \in [0, \pi]$. On the other hand, $\theta^{\dagger}$ can diverge due to the analytic extension of the frequency, $f^\dagger = k/\theta^{\dagger} \to 0$. In linear inflation geometry, we expect that $S(f) \sim f^{-1.5}$   would still hold at $p_{\rm c}$ and the associated length scale perpendicular to the inflation direction would diverge without the need of analytic extension.

{\bf Discussion.}
%In summary, we employ two methods to evaluate the correlation angle of the displacement field $d_{\rm r}(\theta)$: $\theta^*$ determined directly from the correlation function $G(\theta)$ and $\theta^\dagger \equiv k/f^{\dagger}$ based on the inverse Fourier transform of the power spectrum $S(f) = \mathcal{F}[G(\theta)]$. The two angles $\theta^*$ and $\theta^\dagger$ coincide at low $p$.  Near $p_{\rm c}$, $\theta^*$  remains finite due to the cutoff introduced by the finite period  $[0, 2 \pi]$, while $\theta^\dagger$ diverges following a power-law due to the vanishing $f^\dagger$ revealed by the analytic extension of the frequency to zero. 
The angular behavior  of the particle displacements $d_r(\theta)$ in response to the central inflation can be summarized as follows. (i) In the unjammed phase ($p=0$), the displacements are uncorrelated and white-noiselike, with a zero correlation angle. (ii) In the plastic phase ($0<p<p_{\rm c}$),  the correlation angle is finite, and the displacements are correlated only within the correlation angle. (iii)  In the quasielastic phase ($p>p_{\rm c}$), the correlation angle is finite  because the displacement field is dominated by dipolar and quadrupolar patterns.
The long-range angular correlation emerges at the PE transition ($p_{\rm c}$), captured by a diverging correlation angle $\theta^\dagger$ and a power-law spectrum  $S(f)  \sim f^{-1.5}$.

This study opens several questions for future studies. In particular, the origin of the $f^{-1.5}$ spectrum at $p_{\rm c}$ needs to be recovered. 
It  will be useful to generalize the current setup of circular inflation to other types of deformations~\cite{nampoothiri2022tensor,livne2024continuum, kaur2024selection}. 
Despite the similarity between the current problem and two-dimensional  melting, unlike the latter, we expect that the two-step transitions discussed here would also appear in higher dimensions~\cite{charan2023anomalous}.
Finally, it would be very interesting to reconcile the MFRT and MRST approaches. 

%A heuristic argument shows that the  hexatic phase transition in two-dimensional melting is a consequence of the competition between the vortex entropy $TS$ and energy $E$, both of which depend logarithmically on the system size in two dimensions. This argument can not be applied to the current granular system, since the temperature is zero. On the other hand, the match between the overlapping effects due to over-compression and the energy barriers required for plastic events (as discussed in the Introduction) may provide an alternative mechanism universal in any dimensions. 

{\bf Acknowledgments.}
We warmly thank Matteo Baggioli, Bulbul Chakraborty, Wouter Ellenbroek, Wenyue Fan, Michael Moshe, Jin Shang, and Jie Zhang for inspiring discussions. 
We acknowledge financial support from NSFC (Grants 12161141007, 11935002, 12047503 and 12404290), from 
Chinese Academy of Sciences (Grant ZDBS-LY-7017), 
and from Wenzhou Institute (Grant WIUCASQD2023009). 
IP acknowledges support from the ISF under grant \#3492/21 (collaboration with China) and the Minerva Center for ``Aging, from physical materials to human tissues" at the Weizmann Institute. 
In this work access was granted to the High-Performance Computing Cluster of Institute of Theoretical Physics - the Chinese Academy of Sciences.\\

\bibliographystyle{unsrt}
\bibliography{apssamp.bib}

\begin{thebibliography}{10}

\bibitem{liu2010jamming}
Andrea~J Liu and Sidney~R Nagel.
\newblock The jamming transition and the marginally jammed solid.
\newblock {\em Annu. Rev. Condens. Matter Phys.}, 1(1):347--369, 2010.

\bibitem{muller2015marginal}
Markus M{\"u}ller and Matthieu Wyart.
\newblock Marginal stability in structural, spin, and electron glasses.
\newblock {\em Annu. Rev. Condens. Matter Phys.}, 6(1):177--200, 2015.

\bibitem{charbonneau2014fractal}
Patrick Charbonneau, Jorge Kurchan, Giorgio Parisi, Pierfrancesco Urbani, and Francesco Zamponi.
\newblock Fractal free energy landscapes in structural glasses.
\newblock {\em Nature communications}, 5(1):3725, 2014.

\bibitem{hentschel2011athermal}
H.~G.~E. Hentschel, Smarajit Karmakar, Edan Lerner, and Itamar Procaccia.
\newblock Do athermal amorphous solids exist?
\newblock {\em Phys. Rev. E}, 83:061101, Jun 2011.

\bibitem{biroli2016breakdown}
Giulio Biroli and Pierfrancesco Urbani.
\newblock Breakdown of elasticity in amorphous solids.
\newblock {\em Nature physics}, 12(12):1130--1133, 2016.

\bibitem{karmakar2010statistical}
Smarajit Karmakar, Edan Lerner, and Itamar Procaccia.
\newblock Statistical physics of the yielding transition in amorphous solids.
\newblock {\em Phys. Rev. E}, 82:055103, Nov 2010.

\bibitem{morse2020differences}
Peter Morse, Sven Wijtmans, Merlijn Van~Deen, Martin Van~Hecke, and M~Lisa Manning.
\newblock Differences in plasticity between hard and soft spheres.
\newblock {\em Physical Review Research}, 2(2):023179, 2020.

\bibitem{parisi2020theory}
Giorgio Parisi, Pierfrancesco Urbani, and Francesco Zamponi.
\newblock {\em Theory of simple glasses: exact solutions in infinite dimensions}.
\newblock Cambridge University Press, 2020.

\bibitem{charbonneau2014exact}
Patrick Charbonneau, Jorge Kurchan, Giorgio Parisi, Pierfrancesco Urbani, and Francesco Zamponi.
\newblock Exact theory of dense amorphous hard spheres in high dimension. iii. the full replica symmetry breaking solution.
\newblock {\em Journal of Statistical Mechanics: Theory and Experiment}, 2014(10):P10009, 2014.

\bibitem{charbonneau2015numerical}
Patrick Charbonneau, Yuliang Jin, Giorgio Parisi, Corrado Rainone, Beatriz Seoane, and Francesco Zamponi.
\newblock Numerical detection of the gardner transition in a mean-field glass former.
\newblock {\em Physical Review E}, 92(1):012316, 2015.

\bibitem{berthier2016growing}
Ludovic Berthier, Patrick Charbonneau, Yuliang Jin, Giorgio Parisi, Beatriz Seoane, and Francesco Zamponi.
\newblock Growing timescales and lengthscales characterizing vibrations of amorphous solids.
\newblock {\em Proceedings of the National Academy of Sciences}, 113(30):8397--8401, 2016.

\bibitem{urbani2023gardner}
Pierfrancesco Urbani, Yuliang Jin, and Hajime Yoshino.
\newblock The gardner glass.
\newblock In {\em Spin Glass Theory and Far Beyond: Replica Symmetry Breaking After 40 Years}, pages 219--238. World Scientific, 2023.

\bibitem{yoshino2014shear}
Hajime Yoshino and Francesco Zamponi.
\newblock Shear modulus of glasses: Results from the full replica-symmetry-breaking solution.
\newblock {\em Physical Review E}, 90(2):022302, 2014.

\bibitem{jin2017exploring}
Yuliang Jin and Hajime Yoshino.
\newblock Exploring the complex free-energy landscape of the simplest glass by rheology.
\newblock {\em Nature communications}, 8(1):14935, 2017.

\bibitem{jin2018stability}
Yuliang Jin, Pierfrancesco Urbani, Francesco Zamponi, and Hajime Yoshino.
\newblock A stability-reversibility map unifies elasticity, plasticity, yielding, and jamming in hard sphere glasses.
\newblock {\em Science advances}, 4(12):eaat6387, 2018.

\bibitem{nakayama2016protocol}
Daijyu Nakayama, Hajime Yoshino, and Francesco Zamponi.
\newblock Protocol-dependent shear modulus of amorphous solids.
\newblock {\em Journal of Statistical Mechanics: Theory and Experiment}, 2016(10):104001, 2016.

\bibitem{lemaitre2021anomalous}
Ana{\"e}l Lema{\^\i}tre, Chandana Mondal, Michael Moshe, Itamar Procaccia, Saikat Roy, and Keren Screiber-Re'em.
\newblock Anomalous elasticity and plastic screening in amorphous solids.
\newblock {\em Physical Review E}, 104(2):024904, 2021.

\bibitem{mondal2022experimental}
Chandana Mondal, Michael Moshe, Itamar Procaccia, Saikat Roy, Jin Shang, and Jie Zhang.
\newblock Experimental and numerical verification of anomalous screening theory in granular matter.
\newblock {\em Chaos, Solitons \& Fractals}, 164:112609, 2022.

\bibitem{bhowmik2022direct}
Bhanu~Prasad Bhowmik, Michael Moshe, and Itamar Procaccia.
\newblock Direct measurement of dipoles in anomalous elasticity of amorphous solids.
\newblock {\em Physical Review E}, 105(4):L043001, 2022.

\bibitem{kumar2022anomalous}
Avanish Kumar, Michael Moshe, Itamar Procaccia, and Murari Singh.
\newblock Anomalous elasticity in classical glass formers.
\newblock {\em Physical Review E}, 106(1):015001, 2022.

\bibitem{charan2023anomalous}
Harish Charan, Michael Moshe, and Itamar Procaccia.
\newblock Anomalous elasticity and emergent dipole screening in three-dimensional amorphous solids.
\newblock {\em Physical Review E}, 107(5):055005, 2023.

\bibitem{kumar2024elasticity}
Avanish Kumar and Itamar Procaccia.
\newblock Elasticity, plasticity and screening in amorphous solids: A short review.
\newblock {\em Europhysics Letters}, 145(2):26002, feb 2024.

\bibitem{jin2023intermediate}
Yuliang Jin, Itamar Procaccia, and Tuhin Samanta.
\newblock Intermediate phase between jammed and unjammed amorphous solids.
\newblock {\em Phys. Rev. E}, 109:014902, Jan 2024.

\bibitem{van2014contact}
Merlijn~S van Deen, Johannes Simon, Zorana Zeravcic, Simon Dagois-Bohy, Brian~P Tighe, and Martin van Hecke.
\newblock Contact changes near jamming.
\newblock {\em Physical Review E}, 90(2):020202, 2014.

\bibitem{van2016contact}
Merlijn~S van Deen, Brian~P Tighe, and Martin van Hecke.
\newblock Contact changes of sheared systems: Scaling, correlations, and mechanisms.
\newblock {\em Physical Review E}, 94(6):062905, 2016.

\bibitem{fu2024odd}
Yang Fu, H.~George~E. Hentschel, Pawandeep Kaur, Avanish Kumar, and Itamar Procaccia.
\newblock Odd dipole screening in radial inflation.
\newblock {\em Phys. Rev. E}, 110:065003, Dec 2024.

\bibitem{Kumar_2023}
Avanish Kumar, Itamar Procaccia, and Murari Singh.
\newblock Disorder-induced mode coupling and symmetry breaking in amorphous solids.
\newblock {\em Europhysics Letters}, 142(3):36001, apr 2023.

\bibitem{fu2025SI}
See supplemental material for additional simulation data and analyses.

\bibitem{ellenbroek2006critical}
Wouter~G Ellenbroek, Ell{\'a}k Somfai, Martin van Hecke, and WIM vAN SAARLoos.
\newblock Critical scaling in linear response of frictionless granular packings near jamming.
\newblock {\em Physical Review Letters}, 97(25):258001, 2006.

\bibitem{ellenbroek2009jammed}
Wouter~G Ellenbroek, Martin Van~Hecke, and Wim Van~Saarloos.
\newblock Jammed frictionless disks: Connecting local and global response.
\newblock {\em Physical Review E—Statistical, Nonlinear, and Soft Matter Physics}, 80(6):061307, 2009.

\bibitem{makse2000packing}
Hern{\'a}n~A Makse, David~L Johnson, and Lawrence~M Schwartz.
\newblock Packing of compressible granular materials.
\newblock {\em Physical Review Letters}, 84(18):4160, 2000.

\bibitem{o2003jamming}
Corey~S O’hern, Leonardo~E Silbert, Andrea~J Liu, and Sidney~R Nagel.
\newblock Jamming at zero temperature and zero applied stress: The epitome of disorder.
\newblock {\em Physical Review E}, 68(1):011306, 2003.

\bibitem{olsson2007critical}
Peter Olsson and Stephen Teitel.
\newblock Critical scaling of shear viscosity at the jamming transition.
\newblock {\em Physical Review Letters}, 99(17):178001, 2007.

\bibitem{heussinger2009jamming}
Claus Heussinger and Jean-Louis Barrat.
\newblock Jamming transition as probed by quasistatic shear flow.
\newblock {\em Physical Review Letters}, 102(21):218303, 2009.

\bibitem{van2009jamming}
Martin van Hecke.
\newblock Jamming of soft particles: geometry, mechanics, scaling and isostaticity.
\newblock {\em Journal of Physics: Condensed Matter}, 22(3):033101, 2009.

\bibitem{behringer2018physics}
Robert~P Behringer and Bulbul Chakraborty.
\newblock The physics of jamming for granular materials: a review.
\newblock {\em Reports on Progress in Physics}, 82(1):012601, 2018.

\bibitem{pan2023review}
Deng Pan, Yinqiao Wang, Hajime Yoshino, Jie Zhang, and Yuliang Jin.
\newblock A review on shear jamming.
\newblock {\em Physics Reports}, 1038:1--18, 2023.

\bibitem{o2002random}
Corey~S O'Hern, Stephen~A Langer, Andrea~J Liu, and Sidney~R Nagel.
\newblock Random packings of frictionless particles.
\newblock {\em Physical Review Letters}, 88(7):075507, 2002.

\bibitem{wyart2005geometric}
Matthieu Wyart, Sidney~R Nagel, and Thomas~A Witten.
\newblock Geometric origin of excess low-frequency vibrational modes in weakly connected amorphous solids.
\newblock {\em Europhysics Letters}, 72(3):486, 2005.

\bibitem{kubo2012statistical}
Ryogo Kubo, Morikazu Toda, and Natsuki Hashitsume.
\newblock {\em Statistical physics II: nonequilibrium statistical mechanics}, volume~31.
\newblock Springer Science \& Business Media, 2012.

\bibitem{nampoothiri2022tensor}
Jishnu~N Nampoothiri, Michael D'Eon, Kabir Ramola, Bulbul Chakraborty, and Subhro Bhattacharjee.
\newblock Tensor electromagnetism and emergent elasticity in jammed solids.
\newblock {\em Physical Review E}, 106(6):065004, 2022.

\bibitem{livne2024continuum}
Noemie~S Livne, Tuhin Samanta, Amit Schiller, Itamar Procaccia, and Michael Moshe.
\newblock Continuum mechanics of differential growth in disordered granular matter.
\newblock {\em arXiv preprint arXiv:2408.13086}, 2024.

\bibitem{kaur2024selection}
Pawandeep Kaur, Itamar Procaccia, and Tuhin Samanta.
\newblock Selection principle for the screening parameters in the mechanical response of amorphous solids.
\newblock {\em Phys. Rev. E}, 111:015506, Jan 2025.

\bibitem{bitzek2006structural}
Erik Bitzek, Pekka Koskinen, Franz G{\"a}hler, Michael Moseler, and Peter Gumbsch.
\newblock Structural relaxation made simple.
\newblock {\em Physical Review Letters}, 97(17):170201, 2006.

\bibitem{THOMPSON2022108171}
Aidan~P. Thompson, H.~Metin Aktulga, Richard Berger, Dan~S. Bolintineanu, W.~Michael Brown, Paul~S. Crozier, Pieter~J. {in 't Veld}, Axel Kohlmeyer, Stan~G. Moore, Trung~Dac Nguyen, Ray Shan, Mark~J. Stevens, Julien Tranchida, Christian Trott, and Steven~J. Plimpton.
\newblock Lammps - a flexible simulation tool for particle-based materials modeling at the atomic, meso, and continuum scales.
\newblock {\em Computer Physics Communications}, 271:108171, 2022.

\bibitem{goodrich2012finite}
Carl~P Goodrich, Andrea~J Liu, and Sidney~R Nagel.
\newblock Finite-size scaling at the jamming transition.
\newblock {\em Physical Review Letters}, 109(9):095704, 2012.

\bibitem{goodrich2016scaling}
Carl~P Goodrich, Andrea~J Liu, and James~P Sethna.
\newblock Scaling ansatz for the jamming transition.
\newblock {\em Proceedings of the National Academy of Sciences}, 113(35):9745--9750, 2016.

\bibitem{coulais2014shear}
Corentin Coulais, Antoine Seguin, and Olivier Dauchot.
\newblock Shear modulus and dilatancy softening in granular packings above jamming.
\newblock {\em Physical Review Letters}, 113(19):198001, 2014.

\end{thebibliography}

\clearpage
\centerline{\bfseries \bf \Large End Matter}
{\it{Appendix A: Model\textemdash}}
The two-dimensional model employed in our numerical simulations  consists of $N$ equimolar frictionless bidisperse disks with radii $R_{1}=0.5$ and $R_{2}=0.7$ (the length unit is the diameter of small disks).
The normal force $\vec{F}_{ij}$ between particles $i$ and $j$ is
\begin{align}
\vec{F}_{ij}^{(n)} = k_n \Delta_{ij}^{(n)} \hat{\vec{n}}_{ij} - \gamma_{n} \vec{v}_{ij}^{(n)},
\end{align}
where
 $\hat{\vec{n}}_{ij} = \vec{r}_{ij}/r_{ij}$ is the unit interparticle distance, and $\vec{v}_{ij}^{(n)}$ is the normal component of the relative velocity.
The overlap between two contacting particles is defined as $\Delta_{ij}^{(n)}=R_{i}+R_{j}-\left| \bf{r}_{i}-\bf{r}_{j}\right|$.
The spring coefficient is $k_n=k_n^{'} \sqrt{\Delta_{ij}^{(n)} R_{ij}}$ for the Hertzian interaction,
and $k_n=k_n^{'}$
for the Hookean interaction, with $k_n^{'} = 2\times 10^{5}$ and 
$R_{ij}^{-1}=R_{i}^{-1}+R_{j}^{-1}$. 
The damping coefficient is $\gamma_n = 500$ for both models.
The mass of all particles is set to be 1.
The presented data are for the Hertzian potential unless otherwise specified. 
\\
%varied from $N\approx 10^3$ to $10^5$.}\\
%\YF{4016, 16450, and 67352 particles within the fixed boundary, 5548, 22184, 88776 particles for the whole box}\\

{\it{Appendix B: Simulation methods\textemdash}}
The initially random configuration is generated under periodic boundary conditions at a fixed packing fraction $\varphi$. The system is then rapidly quenched to reach mechanical equilibrium, and the mechanical pressure $p$ is measured. 
After this step, simulations are performed in a circular area with a fixed boundary condition: a small particle is randomly chosen as the inflation center at $r=0$, and all particles at $r \geq r_{\rm out}$ are fixed.

The inflation simulation is carried out as follows. We first instantaneously inflate a small disk 
at the center by a factor of 1.2, 
i.e., $r_{\rm in} \to r_{\rm in} + d_0$ with $r_{\rm in} = 0.5$, then minimize the energy of the entire system after the inflation.
The inflation parameter $d_0 = 0.1$ unless otherwise specified. 
The position of the central particle is fixed during the minimization.
For jammed configurations with $p>0$ and $\varphi > \varphi_{\rm J}$,  the energy is dissipated by damped molecular dynamics (MD) simulations following the previous study~\cite{jin2023intermediate}. 
%with a viscoelastic damping factor $\eta_n = 500$. 
The minimization procedure is terminated when the net force per particle $F_{\rm net} \leq 10^{-7}$. 
For unjammed configurations ($p=0$, $\varphi < \varphi_{\rm J}$), the damped MD simulation can create an inhomogeneous flow, which results in large voids that are empty of particles. 
To avoid such highly inhomogeneous displacements, the FIRE algorithm~\cite{bitzek2006structural}, which minimizes the energy more rapidly  than the damped MD algorithm, is used for unjammed configurations.
The above protocols are implemented using the LAMMPS package \cite{THOMPSON2022108171}.
The presented data are averaged over 
%more than \YF{for} 
400 independent samples.
{Rattlers (particles with fewer than $d+1$ contacts) are removed from the mechanical equilibrium  configurations  before and after the inflation.}\\

\begin{figure}[htbp]
  \centering
\includegraphics[width=0.9\linewidth]{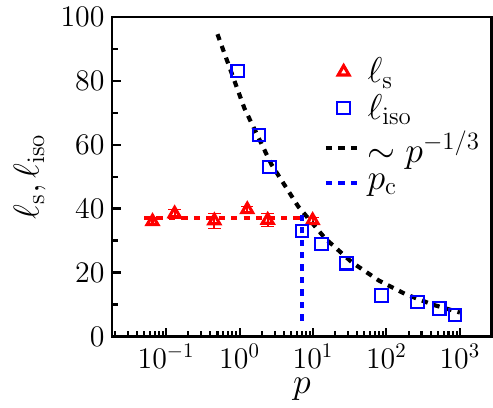}
    \caption{{\bf Matching of the screening length $\ell_{\rm s}$ and the isostatic length $\ell_{\rm iso}$ at $p_{\rm c}$.} The data points of  $\ell_{\rm iso}$ are obtained  in Sec.~S8 of SM ~\cite{fu2025SI},
    %~\YF{\ref{SI: Isostatic length} \cite{fu2025SI}}), 
    and the dashed line represents power-law fitting $\ell_{\rm iso} = 75 p^{-1/3}$. The data points of $\ell_{\rm s}$ are determined by $\ell_{\rm s} =  1/\kappa_{\rm fit}$ using $\kappa_{\rm fit}$ in Fig.~\ref{Fig:2PT}(a), and the horizontal line represents a constant $\ell_{\rm s}\approx 37$.
    %~\YF{$\approx 1/\kappa_1^*\approx \it{r}_{\rm out}/\pi$}. 
    The marked $p_{\rm c} \approx 7$ is estimated independently at the pressure where the power spectrum  is a power-law $S(f) \sim f^{-1.5}$ (see Fig.~\ref{Fig:Correlation_PS}(b)).
    }   
	\label{Fig:matching}
\end{figure}

{\it{Appendix C: Matching isostatic and screening lengths at $p_{\rm c}$\textemdash}}
In Ref.~\cite{jin2023intermediate}, we argue that the isostatic length $\ell_{\rm iso}$ and the screening length $\ell_{\rm s}$ are identical at the PE transition. 
With  $\ell_{\rm s} = 1/\kappa_{\rm fit}$ determined in Fig.~\ref{Fig:2PT} and  $\ell_{\rm iso}$ estimated based on the fluctuations of the relative displacement $d_\parallel$ between contact particles (see details in Sec.~S8 of SM~\cite{fu2025SI}),
%~\YF{\ref{SI: Isostatic length} \cite{fu2025SI}}), 
we can  compare them quantitatively. As expected, the two lengths match  near $p_{\rm c}$ (see Fig.~\ref{Fig:matching}). Note that $p_{\rm c} \approx 7$ is determined independently in Fig.~\ref{Fig:Correlation_PS}(b) using the criterion of a power-law spectrum $S(f) \sim f^{-1.5}$. Figure~\ref{Fig:matching} shows strong evidence that the PE transition results from  the combined effects of plastic screening and isostatic stability.

\begin{figure*}[htbp]
  \centering
\includegraphics[width=0.7\linewidth]
{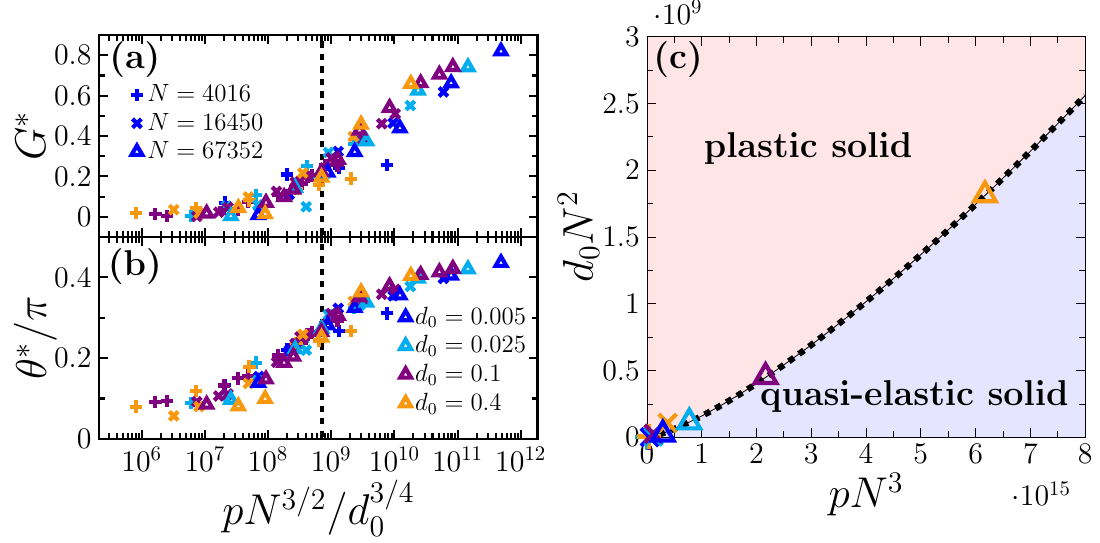}
%{fig6_d0.pdf}
    \caption{{\bf Dependence of the PE transition on the inflation $d_0$ and the system size $N$, for Hertzian systems.}
    (a) $G^*$ and (b) $\theta^*$ as functions of $pN^{3/2}/d_0^{3/4}$, for three different $N$ and four different $d_0$, where the dotted line marks $p_{\rm c} N^{3/2}/d_0^{3/4}$. %~\YF{$p_{\rm c}$ is different for different $N$ and $d_0$, and the dotted line marks the same $p_{\rm c} N^{3/2}/d_0^{3/4}$}. 
    (c) The $N$-rescaled $d_0-p$ phase diagram, where the dotted line is $(N^2 d_0) \sim (N^3 p_{\rm c})^{4/3}$.
    %d0=0.0073*pc^{4/3}
    }   
	\label{Fig:d0}
\end{figure*}

\begin{figure}[htbp]
  \centering
\includegraphics[width=0.9\linewidth]{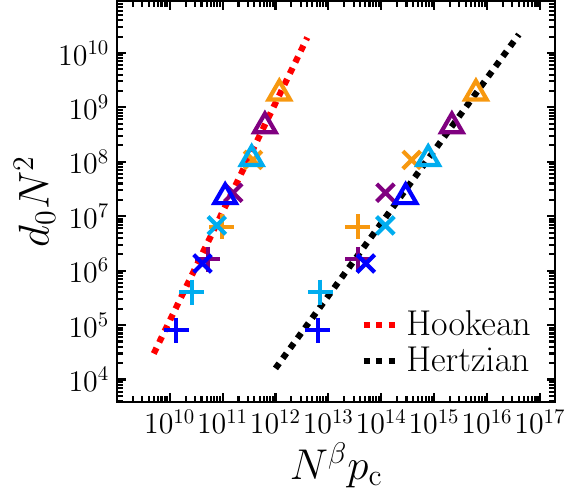}
    \caption{{\bf Power-law scaling between $N^2 d_0$ and $N^\beta p_{\rm c}$.}
    The data are compared to the scaling function Eq.~(\ref{eq:scaling_function2}) for both Hertzian systems, $N^2 d_0 \sim (N^{3} p_{\rm c})^{4/3}$, and Hookean systems, $N^2 d_0 \sim (N^{2} p_{\rm c})^{2}$.
    %Scaling collapse over two decades in the scaled transition pressure $N^{\beta} p_{\rm c}$ and four decades in the scaled level of inflation $d_{0}N^2$. The dotted line distinguishes the plastic and quasielastic regimes above and below, while the red and black dotted lines represent the boundary lines for the Hookean and Hertzian potentials, respectively.
    }
	\label{Fig:scaling}
\end{figure}

{\it Appendix D: Scaling function of $p_{\rm c}(N, d_0)$\textemdash} In the main text, we have shown the $N$ dependence of $p_{\rm c}$, $p_{\rm c} \sim N^{-3/2}$, in Hertzian systems. Intuitively, the PE transition should also depend  on the amount of inflation $d_0$, since larger inflation could introduce stronger plasticity and nonlinear effects. 
%Here we numerically examine the dependence of the PE transition on both $N$ and $d_0$. 
Examining the dependence of the PE transition on the inflation parameter $d_0$ reveals that $p_{\rm c} \sim d_0^{3/4}$ for a fixed $N$, in Hertzian systems.
In Figs.~\ref{Fig:d0}(a) and \ref{Fig:d0}(b), we plot $\theta^*$ and $G^*$ obtained from inflation simulations with a few different $d_0$ and $N$. The data can be collapsed as functions of $pN^{3/2}/d_0^{3/4}$.

{Previous studies~\cite{goodrich2012finite, goodrich2016scaling} have suggested a general jamming scaling  ansatz, $N^\gamma A = \mathcal F(N^\beta p)$, for any relevant physical quantity $A$.
The exponent  $\beta = 2(\alpha-1)$ is related to  the exponent  $\alpha$ in the  interaction potential $U(\delta) \sim \delta^  \alpha$  ($\delta$ is the dimensionless interparticle overlap): $\alpha = 5/2, \beta =3$ for the Hertzian interaction, and  $\alpha = \beta = 2 $ for the Hookean interaction. 
The exponent $\gamma$ is $A$ dependent. 
For example, $\gamma = 1$ in the scaling of the access coordination number $\Delta Z = Z - Z_{\rm iso}$~\cite{goodrich2012finite, goodrich2016scaling}.}

We assume that the inflation threshold $d_{\rm c}(N,p)$ at the PE transition follows the above general scaling ansatz. Then the task is to determine the exponent $\gamma$  and the function form $\mathcal{F}(x)$ for $d_{\rm c}$. Once the scaling function $d_{\rm c}(N,p)$ is known, it can be easily transformed into the scaling function of the transition pressure, $p_{\rm c}(N, d_0)$. Our derivation consists of the following two steps.

First, the exponent $\gamma = 2$ of $d_{\rm c}$ is assumed to be identical to the exponent of the minimum compression strain $\epsilon$ required to break an existing contact or make a new contact. Previous studies~\cite{van2014contact, van2016contact, morse2020differences} suggest that  the scaling function of $\epsilon$ is,
$N^2 \epsilon = \mathcal F(N^\beta p)$, where $\mathcal F(x \ll 1) \sim \rm{constant}$, $\mathcal F(x \gg 1) \sim x^{1/\beta}$ for the contact making $\epsilon$, and $\mathcal F(x \ll 1) \sim x$, $\mathcal F(x \gg 1) \sim x^{1/\beta}$ for the contact breaking $\epsilon$.
The definitions of $\epsilon$ and $d_{\rm c}$ are different: $\epsilon$ is the microscopic strain separating the rigorous elastic regime without any contact changing and the plastic regime; in contrast, $d_{\rm c}$ is the deformation separating the quasielastic regime, where many contact changes can occur but the overall response can be coarse-grained into elasticlike behavior (see Sec.~S9 of SM~\cite{fu2025SI} 
%~\YF{\ref{SI: Contact changes} \cite{fu2025SI}} 
for contact changes in our simulations),  and the anomalous regime dominated by plasticity. However, it can be seen that $\epsilon$ and $d_{\rm c}$ have analogous meanings, and thus one may expect 
the same exponent $\gamma=2$ for both parameters. This expectation is validated by the simulation data, as discussed below.

{
Second, at the PE transition, the isostatic length $\ell_{\rm iso} \sim \Delta Z^{-1} \sim p^{-1/\beta}$  
and the screening length $\ell_{\rm s} \sim N^{1/2}$ (in two dimensions) should be matched (see Appendix C).
%, as argued in Ref.~\cite{jin2023intermediate} and the main text. 
This gives the scaling $p_{\rm c} \sim N^{-\beta/2}$.  
Imposing this scaling to the general scaling ansatz of $d_{\rm c}$,  we obtain a scaling function,
\begin{equation}
N^2 d_{\rm c}(N,p) = \mathcal{F}(N^\beta p),
\label{eq:scaling_function1}
\end{equation}
with $\mathcal{F}(x) \sim x^{4/\beta}$. Taking $d_{\rm 0}$ as the independent control parameter, Eq.~(\ref{eq:scaling_function1}) can be alternatively written as,
\begin{equation}
N^\beta p_{\rm c}(N, d_0) = \mathcal{G}(N^2 d_0),
\label{eq:scaling_function2}
\end{equation}
with $\mathcal{G}(x) \sim x^{\beta/4}$. 
Equation~(\ref{eq:scaling_function2}) is confirmed by our simulation data in Fig.~\ref{Fig:scaling} for both Hertzian and Hookean systems.
The scaling relationship $p_{\rm c} \sim N^{-\beta/2} d_0^{\beta/4}$ also explains the data collapsing in Fig.~\ref{Fig:d0}.  
}

Based on this analysis, a $N$-rescaled $d_0-p$  phase diagram is proposed (see Fig.~\ref{Fig:d0}(c)). The boundary between quasielastic and plastic regimes is $N^2 d_0 \sim (N^\beta p_{\rm c})^{4/\beta}$. 
Additional data for Hookean disks are reported in SM~\cite{fu2025SI}.
Interestingly, a similar strain-pressure phase diagram is conjectured in Ref.~\cite{coulais2014shear} (the two regimes are called linear and shear softening regimes there), although the linear regime could not be accessed in that experiment.

\onecolumngrid

\centerline{\bf \Large Supplementary Information}
\vspace{1cm}

%\centerline{\bf \Large Supplementary Material}
\tableofcontents
\setcounter{figure}{0}
\setcounter{equation}{0}
\setcounter{table}{0}
\renewcommand\thefigure{S\arabic{figure}}
\renewcommand\theequation{S\arabic{equation}}
\renewcommand\thesection{S\arabic{section}}
\renewcommand\thetable{S\arabic{table}}

\clearpage

%S1
\section{Dipolar and quadrupolar patterns of the displacement fields at $p>p_{\rm c}$ under square periodic boundary conditions}
\label{SI: Dipolar and quadrupolar}

\begin{figure}[htbp]
  \centering
\includegraphics[width=0.8\linewidth]{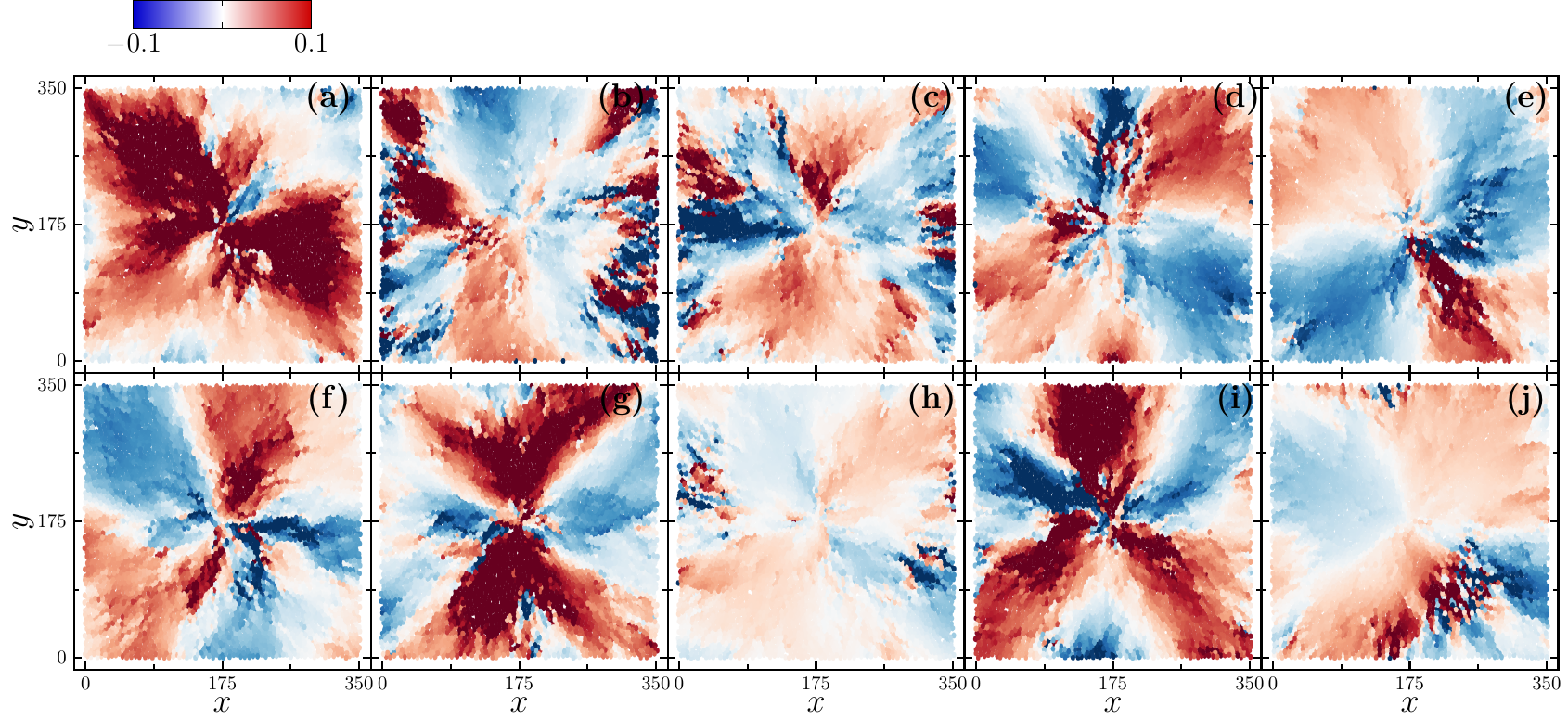}
    \caption{{\bf Radial displacement fields under periodic boundary conditions.}
    Presented in (a-j) are the heap maps of $\hat{d}_r(x,y)$  obtained for ten independent samples of $N=2774$ disks at $p = 3.5 \times 10^3$.
    }
	\label{Fig:rec_PBC_displacement}
\end{figure}

In the quasi-elastic regime ($p>p_{\rm c}$), dipolar and quadrupolar patterns of the displacement fields are commonly observed (see Fig.~1(\textit{g,j})). To examine if such anisotropic patterns are caused by the fixed circular boundary conditions,  we  perform additional inflation simulations in a square simulation box with periodic boundary conditions (the center of mass of the entire system is fixed). As shown in Fig.~\ref{Fig:rec_PBC_displacement}, dipolar and quadrupolar patterns are repeatedly observed in different samples, confirming that the anisotropic patterns at $p> p_{\rm c}$ do not significantly depend on the boundary conditions. \\

\clearpage

%S2

\section{Ensemble-averaged radial displacements}
\label{SI: Ensemble-averaged radial displacements}

\begin{figure}[htbp]
  \centering
\includegraphics[width=0.8\linewidth]{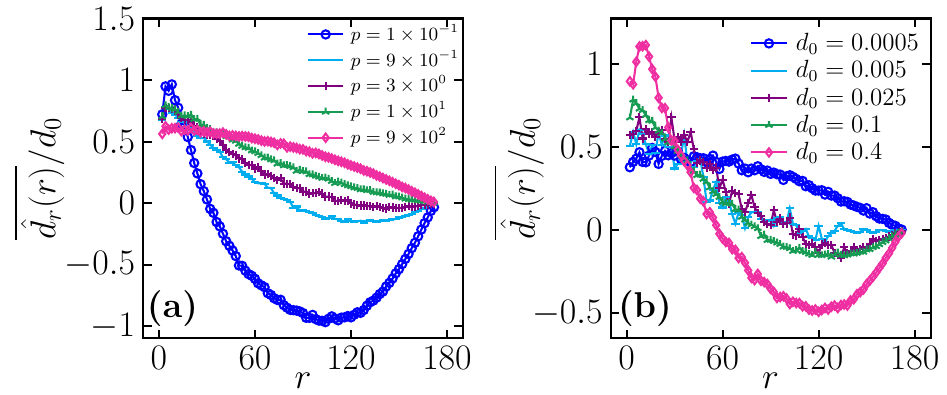}
    \caption{{\bf Ensemble-averaged $\overline{\hat{d}_r(r)}$ for $N=67352$.}
    (a) A few different pressures $p$, for a fixed $d_0 = 0.1$. (b) A few different $d_0$, for a fixed $p=9\times10^{-1}$.
    }
	\label{Fig:r_dr_averge_max}
\end{figure}

The data of rescaled radial displacements $\hat{d}_r(r)$ presented in Fig.~1(\textit{b,e,h,k}) are obtained for single configurations. The single-sample curves are suitable to fit the solutions of the mechanical response screening theory (MRST). Previous inflation simulations have also considered ensemble-averaged quantities, which are averaged over many independent configuration samples~\cite{ellenbroek2006critical, ellenbroek2009jammed}. The ensemble-averaged $\overline{\hat{d}_r(r)}$ are presented in Fig.~\ref{Fig:r_dr_averge_max}, for a few different pressures $p$ and inflation parameters $d_0$. 
For the largest $p$ and smallest $d_0$, the behavior of $\overline{\hat{d}_r(r)}$ is close to the elastic solution Eq.~(1). 
With decreasing $p$ or increasing $d_0$, deviations from Eq.~(1) are observed. 
When $p$ is near or below $p_{\rm c}$, or when $d_0$ is near or larger than $d_{\rm c}$, the  ensemble-averaged $\overline{\hat{d}_r(r)}$  mix samples with elastic and anomalous responses, and thus they can be fitted by neither Eq.~(1) nor Eq.~(3). 
In order to compare with theoretical solutions (Eq.~(1) or (3)), one should use single-sample data  $\hat{d}_r(r)$ instead of the ensemble-averaged data $\overline{\hat{d}_r(r)}$. Fig.~\ref{Fig:r_dr_averge_max} also shows that the coarse-grained elasticity ~\cite{ellenbroek2006critical, ellenbroek2009jammed} works in the quasi-elastic phase ($p>p_{\rm c}$ and $d_0<d_{\rm c}$), but fails in the plastic phase ($p<p_{\rm c}$ and $d_0>d_{\rm c}$) where nonlinear behavior is important. \\

\clearpage

%S3
\section{Jamming transition}
\label{SI: Jamming transition}

\begin{figure}[htbp]
  \centering
\includegraphics[width=0.7\linewidth]{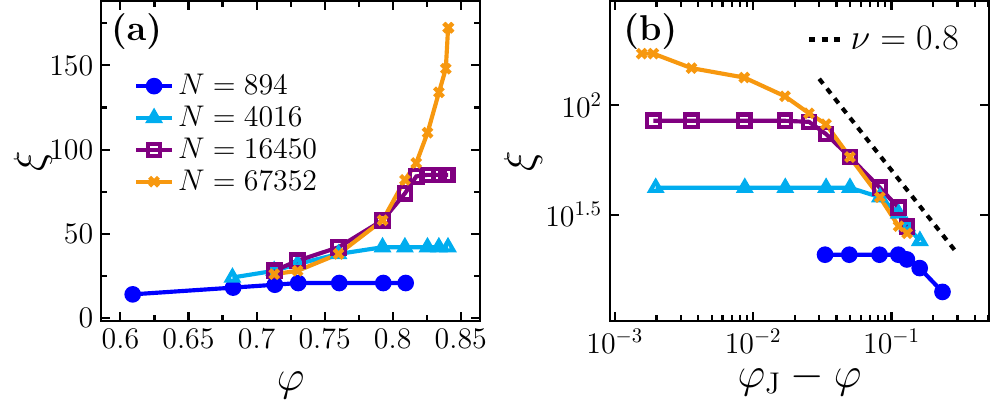}
    \caption{{\bf Jamming transition}.
    The characteristic length $\xi$ for the jamming transition as a function of (a) the packing density $\varphi$, and (b) $\varphi_{\rm J} - \varphi$, for a few different $N$. 
    }   
	\label{Fig:jamming_transtion}
\end{figure}

\begin{figure}[htbp]
  \centering
\includegraphics[width=1\linewidth]{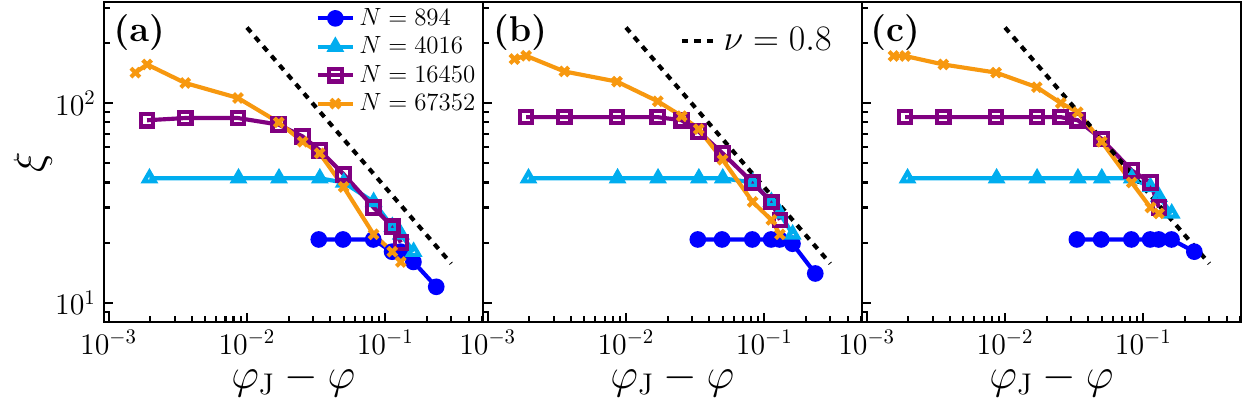}
    \caption{{\bf  Correlation length $\xi$ obtained using different thresholds.} The threshold  $d_{\rm th}$ in the definition $d_r(r=\xi) = d_{\rm th}$ is respectively (a) $10^{-4}$, (b) $10^{-5}$ and (c)  $10^{-6}$.
    The exponent $\nu \approx 0.8$ is unchanged. 
    }
	\label{Fig:unjam_dr_1e-3-6}
\end{figure}

The unjammed configurations with $p=0$ are studied and it is found that the displacement response of the central inflation propagates only up to a finite distance. 
Without loss of generality, we define the length scale $\xi$ by $d_r(r=\xi) = d_{\rm th} = 10^{-3}$.
The power-law divergence Eq.~(3) of $\xi$ with $\nu \approx 0.8$ is confirmed in Fig.~\ref{Fig:jamming_transtion} when approaching $\varphi_{\rm J} = 0.842$, consistent with previous studies ~\cite{o2002random, heussinger2009jamming}.
The exponent $\nu$ is unchanged when the threshold $ d_{\rm th}$ in the definition of $\xi$  is decreased from $10^{-3}$ to $10^{-6}$ (see Fig.~\ref{Fig:unjam_dr_1e-3-6}).

\clearpage

%S4
\section{Dependence of $\theta^*$ on the threshold $G_{\rm th}$}
\label{SI: Dependence on the threshold}

\begin{figure}[htbp]
  \centering
\includegraphics[width=1\linewidth]{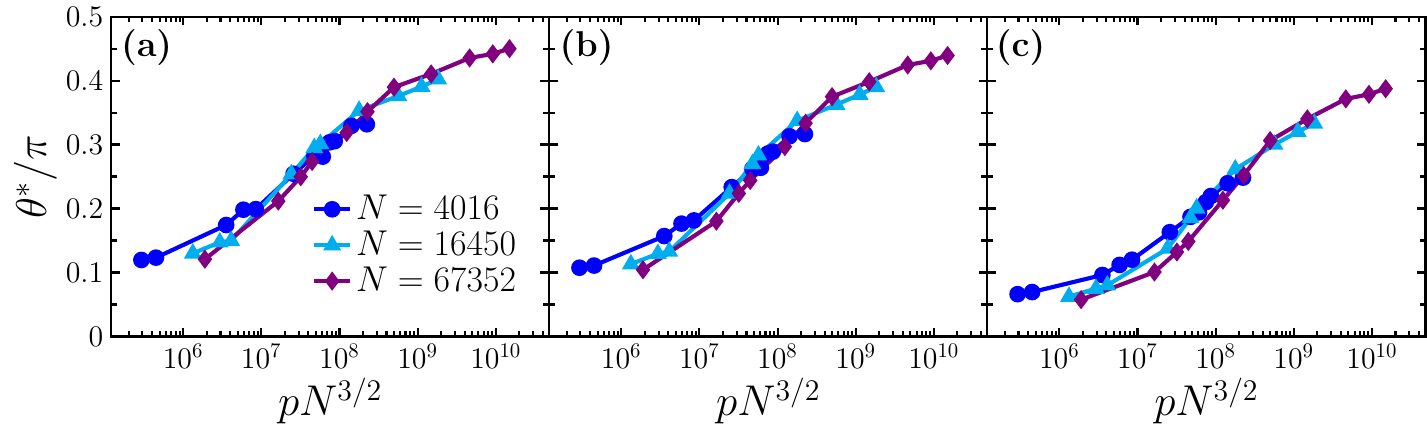}
    \caption{{\bf Correlation angle $\theta^{*}$ as a function of $pN^{3/2}$, using different threshold values $G_{\rm th}$.}
    In (a-c), $G_{\rm th} = 0.02, 0.05, 0.2$ respectively.
    %Dependence of correlation angle $\theta^{*}$ on the threshold $G_{\rm th}$.}$\theta^*$ defined by $G(\theta=\theta^\ast) = G_{\rm th}$ as a function of pressure for (a)  $N=4016$, (b)  $N=16450$, and (a)  $N=67352$. 
    }
	\label{Fig:theta_star_shift}
\end{figure}

%In the main text, we have shown the $N$-dependence of $p_{\rm c}$, $p_{\rm c} \sim N^{-3/2}$, in Hertzian systems. 
Figure~\ref{Fig:theta_star_shift} shows that the finite-size scaling $p_{\rm c} \sim N^{-3/2}$ in Hertzian systems is nearly independent of the threshold value $G_{\rm th}$ used in the definition of $\theta^\ast$, $G(\theta=\theta^\ast) = G_{\rm th}$.

\clearpage

%s5
\section{Additional data on the scaling analysis in Hookean systems}
\label{SI: scaling analysis in Hookean systems}

%Dependence of the plastic-to-elastic (PE) transition pressure $p_{\rm c}$ on the system size $N$ and the  inflation $d_0$ in both Hertzian and Hookean systems}}

\begin{figure}[htbp]
  \centering
\includegraphics[width=0.5\linewidth]{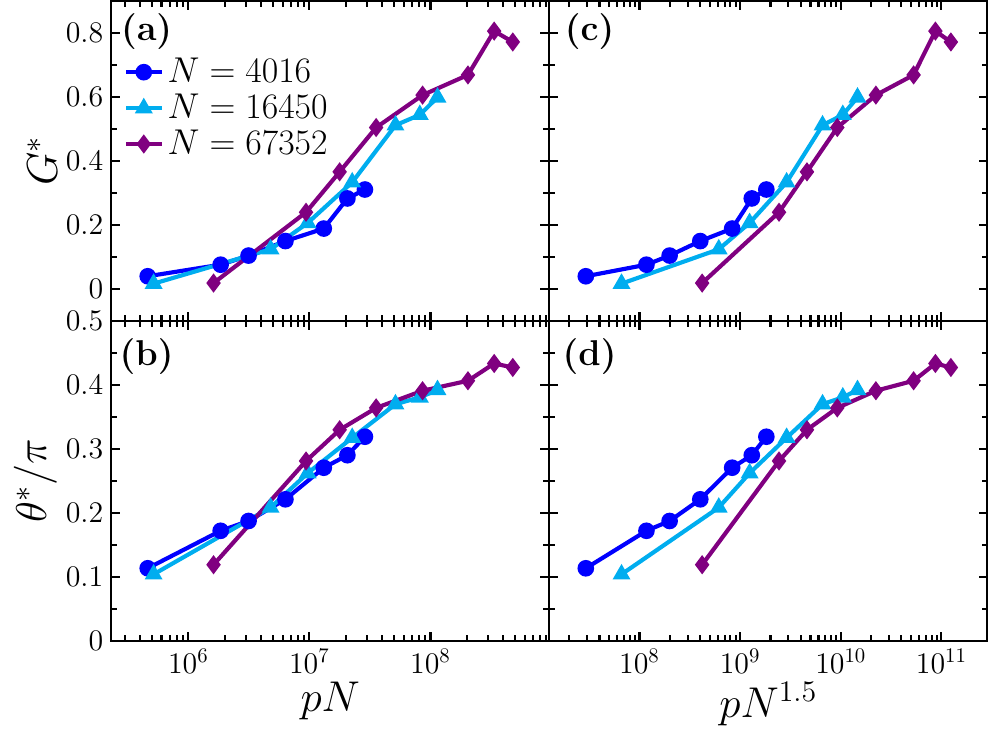}
    \caption{{\bf Finite-size scaling in Hookean disks}.
    The simulation data of $G^*$ and $\theta^*$  can be collapsed as functions of 
    (a,b) $pN$, but not (c,d) $pN^{3/2}$. 
    The inflation $d_0=0.1$ is fixed.
    }
    \label{fig:Hookean}
\end{figure}

\begin{figure*}[htbp]
  \centering
\includegraphics[width=0.7\linewidth]
{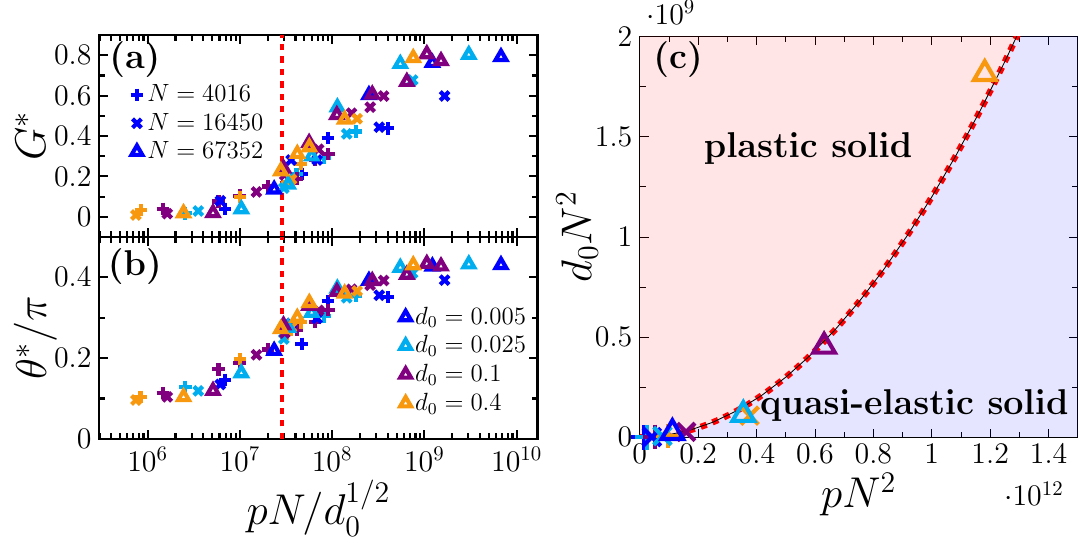}
%{fig6_d0.pdf}
    \caption{{\bf Dependence of the PE transition on the inflation $d_0$ and the system size $N$, for Hookean systems.}
    (a) $G^*$ and (b) $\theta^*$ as functions of $pN/d_0^{1/2}$, for three different $N$ and four different $d_0$, where the dotted line marks $p_{\rm c}N/d_0^{1/2}$. %~\YF{The dotted line marks the same $p_{\rm c} N/d_0^{1/2}$}.
    (c) The $N$-rescaled $d_0-p$ phase diagram, where the dotted line is $(N^2 d_0) \sim (N^2 p_{\rm c})^{2}$. 
    }   
	\label{Fig:d0_Hooke}
\end{figure*}

In Hertzian systems, we have shown the $N$-dependence of $p_{\rm c}$, $p_{\rm c} \sim N^{-3/2}$, in the main text.
For Hookean disks, the numerical results suggest a finite-size scaling $p \sim N^{-1}$, instead of $p \sim N^{-3/2}$, for a fixed $d_0$ (see Fig.~\ref{fig:Hookean}). In addition, the data of $\theta^*$ and $G^*$ can be collapsed as functions of $pN/d_0^{1/2}$ (see Fig.~\ref{Fig:d0_Hooke}).

\clearpage

%S6
\section{Dependence of the power spectrum on the radius}
\label{SI: Dependence of the power spectrum on the radius}

\begin{figure}[htbp]
  \centering
\includegraphics[width=0.8\linewidth]{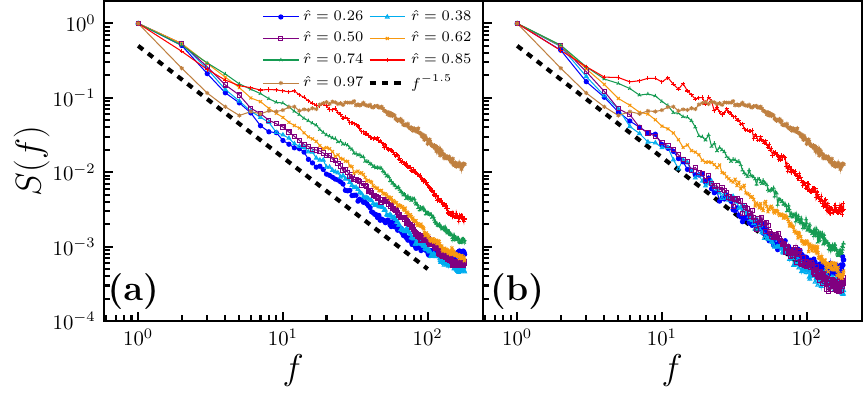}
    \caption{{\bf Power spectrum $S(f)$ at the PE transition $\hat{p}_{\rm c}$ for (a) Hertzian and (b) Hookean potentials, with $N=67352$.} 
    The dotted line represent a power-law decay $f^{-1.5}$.
    }   
	\label{Fig:Correlation_at_pc}
\end{figure}

Fig.~\ref{Fig:Correlation_at_pc} shows that the power-law behavior of the power spectrum   $S(f) \sim f^{-1.5}$  at $\hat{p}_{\rm c}$ is robust against varying $\hat{r}$, although the boundary effects inevitably kink in when $r$ is close to $r_{\rm out}$. Note that the data  collapse for $\hat{r} < 0.6$.
The exponent $1.5$ is observed in both Hertzian and Hookean potentials. 
\\

%We point out that the origin of the $1/f^{1.5}$ noise remains to be explored in future studies. 

\clearpage

%s7
\section{Inverse Fourier transform of the generalized Lorentzian spectrum}
\label{SI: Inverse Fourier transform}

\begin{figure}[htbp]
  \centering
\includegraphics[width=0.5\linewidth]{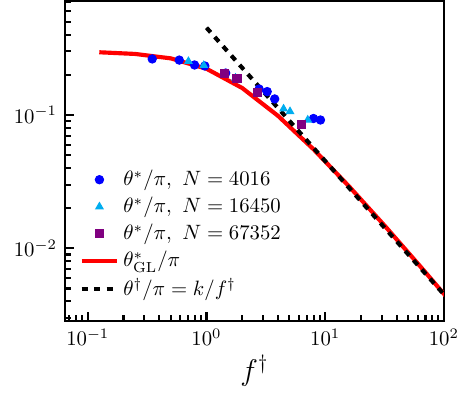}
    \caption{{\bf Correlation angle from the inverse Fourier transform of the generalized Lorentzian spectrum.} 
    $\theta_{\rm GL}^*$ as a function of $f^\dagger$ (red line). Data points are $\theta^*$ obtained from simulations (we only show $\theta^*$ with $\hat{p}<\hat{p}_{\rm c}$). 
    % Three correlation angles change with the characteristic wave number, $k^{\dagger}$, where the angles are measured by the angular correlation function of simulation $\theta^{*}$ (sphere), the correlation angle numerically reproduced from Fourier transform of the theoretical power spectrum $\theta_{\rm num}$ (solid line), and $\theta^{\dagger}=k/f^{\dagger}$ based on Eq.~\ref{eq:Lorentzian} where $k \approx 116$. 
    }
	\label{Fig:Lorentzian}
\end{figure}

According to the Weiner-Khintchine theorem \cite{kubo2012statistical},  the power spectrum $S(f)$ of $d_r(\theta)$ and its auto-correlation $G(\theta)$ form a Fourier transform pair.
The inverse Fourier transform of the  generalized Lorentzian spectrum (Eq.~(5) in the main text) gives,
\begin{align}
G_{\rm GL}(\theta)= \sum_{f = 1}^{\infty} \frac{ B \cos(f \theta)}{1+(f/f^{\dagger})^{1.5}},
\label{eq:GGL}
\end{align}
where $B=\sum_{f} \frac{1}{1+(f/f^{\dagger})^{1.5}}$ is the normalizing factor, and $f$ is an integer because $d_r(\theta)$ is periodic in $\theta \in [0, 2\pi]$.
Following the main text, we define $\theta^*_{\rm GL}$ by $G_{\rm GL}(\theta = \theta^*_{\rm GL} ) = 0.1$.
As shown in Fig.~\ref{Fig:Lorentzian}, the curve $\theta^*_{\rm GL}(f^\dagger)$ is close to the simulation data $\theta^*$ (for $\hat{p}<\hat{p}_{\rm c}$). 
In the large $f^\dagger$ limit (i.e., $\hat{p}\to 0$), $\theta^*_{\rm GL}(f^\dagger \to \infty) = k/f^\dagger$, with $k\approx 0.45$. 
Based on this, we define $\theta^\dagger = k/f^\dagger$ in the main text.
In the small $f^\dagger$ limit (i.e., $\hat{p}\to \hat{p}_{\rm c}$), $\theta^*_{\rm GL}(f^\dagger \to 0) = \theta_{\rm GL,c}^* \approx 0.3 \pi$. In addition, 
$G^*_{\rm GL,c} \equiv - G_{\rm GL}(\theta = \pi ) \approx 0.3$.

According to the above analyses, we conclude that $\theta^*$ and $G^*$  converge to finite values at $\hat{p}_{\rm c}$, $\theta^*_{\rm c} \approx 0.3 \pi$ and $G^*_{\rm c} \approx 0.3$. In other words, even though $S(f) \sim f^{-1.5}$ is clearly scale-free at $\hat{p}_{\rm c}$, its inverse Fourier transform $G(\theta)$ does not show a diverging correlation angle because it has to satisfy the periodic boundary conditions at  $\theta =0$ and $\theta = 2\pi$, such that $f$ needs to be an integer with minimum $f=1$ in Eq.~(\ref{eq:GGL}). This constraint comes from the current setup of circular inflation. If one instead performs  line inflation, then the real variable ($f=n/L$) analytic extension is possible for Fourier modes $\cos\left(\frac{2n \pi x}{L}\right)$, when the linear size of the system $L \to \infty$.

\clearpage

%s8
\section{Isostatic  length}
\label{SI: Isostatic length}

\begin{figure}[htbp]
  \centering
\includegraphics[width=0.9\linewidth]{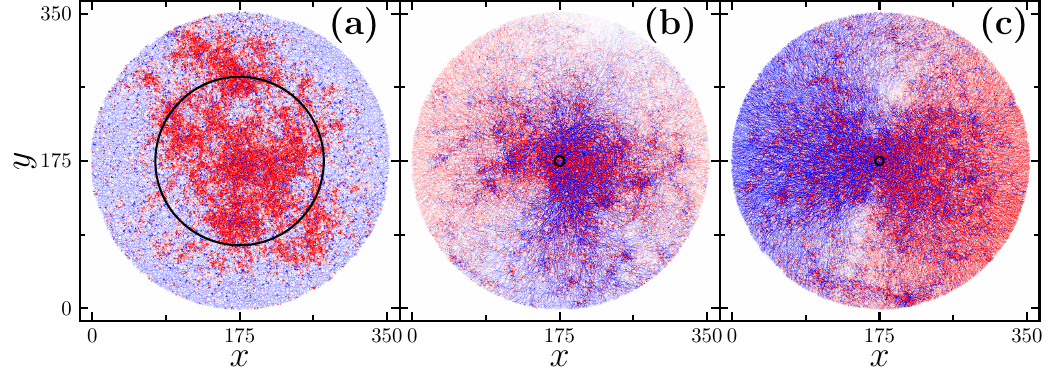}
    \caption{
    {\bf Visualization of the $d_\parallel$ field at (a) $p=4.4 \times 10^{-2}$, (b) $p=5.2 \times 10^2$ and (c) $p=8.6 \times 10^2$, for $N=67452$ and $d_0 = 0.1$.} The red (blue) lines indicate compressed (stretched) bonds with $d_{\parallel}<0$ ($d_{\parallel}>0$). The black circles indicate $\ell_{\rm iso}$.}
	\label{Fig:d_parallel}
\end{figure}

The isostatic length $\ell_{\rm iso}$ has been determined explicitly in previous central inflation simulations, using the fluctuation data of the change in contact force, $\Delta F$~\cite{ellenbroek2006critical, ellenbroek2009jammed}. 
For a pair of contacting particles $i$ and $j$, by definition the force change $\Delta F_{ij}$ is proportional to $d_{\parallel, ij}$, i.e., $\Delta F_{ij} \equiv - k_n d_{\parallel, ij}$, where $d_{\parallel, ij}$ is the parallel component of the relative displacement $\vec{d}_{ij} = \vec{d}_{j}-\vec{d}_{i}$ along the contact bond, and $k_n$ is the spring coefficient.
Inspired by this method, here we use the fluctuations of $d_{\parallel}$ to extract $\ell_{\rm iso}$. Note the difference between $d_r$ and $d_{\parallel}$: $d_r$ is the projection of the displacement  $\vec{d}$ onto the radial direction with respect to the inflation center of the simulation system; $d_{\parallel}$ is the projection of the relative displacement $\vec{d}_{ij}$ onto the bond direction, which represents the change in the bond length. Comparing the visualized fields of $d_{\parallel}(r,\theta)$ in Fig.~\ref{Fig:d_parallel} with the corresponding $d_{r}(r,\theta)$ in Fig.~1(\textit{d, g, j}), one sees that $d_{\parallel}(r,\theta)$ captures the  physics of isostatic stability, rather than that of the plastic screening effects. 
The $d_{\parallel}(r,\theta)$ field looks disordered within a length scale $\ell_{\rm iso}$ from the inflation center. 
At larger length scales beyond  $\ell_{\rm iso}$, the role of disorder is smeared out and the $d_{\parallel}(r,\theta)$ field looks more similar to what is expected for a continuous elastic medium. In other words,  $d_{r}(r,\theta)$ and $d_{\parallel}(r,\theta)$ encode respectively the information of the screening length $\ell_{\rm s}$ and the isostatic length $\ell_{\rm iso}$.

\begin{figure}[htbp]
  \centering
\includegraphics[width=0.8\linewidth]{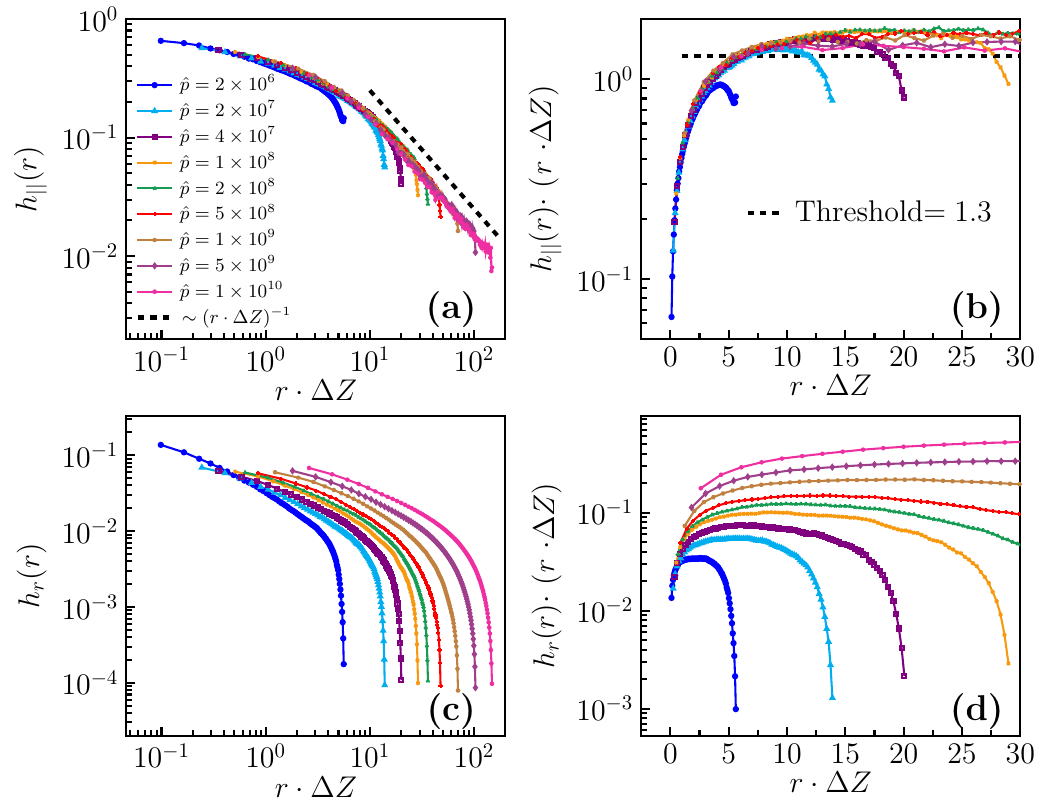}
    \caption{ 
  {\bf Fluctuations of $d_\parallel$.}
(a)  The root mean square fluctuations $h_{\parallel}(r)$ as functions of $r \Delta Z$ for different $p$. 
The deviations from the master curve at large $r$ are due to boundary effects. 
(b)  $h_{\parallel}(r) \times (r \Delta Z)$ as functions of  $r \Delta Z$, showing a plateau at large distances. We choose a threshold $h_{\parallel}(r) \times (r \Delta Z)=1.3$ (dashed line) that gives the crossover value $\ell_{\rm iso} \Delta Z = C \approx 6$.
For comparison, the fluctuations of $d_r(r)$, defined by $h_{r}(r) = \sqrt{\left \langle \left ( d_{r}(r) - \langle d_{r}(r) \rangle \right )^2 \right \rangle}$ are plotted in (c,d), which do not collapse as functions of $r \Delta Z$.
    }   
	\label{Fig:l_iso}
\end{figure}

In order to evaluate $\ell_{\rm iso}$ quantitatively, we follow Refs.~\cite{ellenbroek2006critical, ellenbroek2009jammed} and compute the root mean square fluctuations $h_{\parallel}(r) = \sqrt{\left \langle \left ( d_{\parallel}(r) - \langle d_{\parallel}(r) \rangle \right )^2 \right \rangle}$, where the average $\langle x \rangle$ is taken over along the circle of radius $r$. 
As shown in Fig.~\ref{Fig:l_iso}, the data of $h_{\parallel}(r)$ at different $p$ all collapse when plotted as functions of $r \Delta Z$ with $\Delta Z = Z-Z_{\rm iso}$, suggesting the characteristic length scale $\ell_{\rm iso} \sim 1/\Delta Z$.
The tail of $h_{\parallel}(r)$ follows a power-law $h_{\parallel}(r) \sim r^{-1}$, corresponding to the regime where the effects of disorder disappear. The threshold to reach this asymptotic power-law defines a proper pre-factor for the scaling of  $\ell_{\rm iso}$ (see Fig.~\ref{Fig:l_iso}\textit{b}),
\begin{equation}
\ell_{\rm iso} = \frac{C}{\Delta Z},
\label{eq:l_iso}
\end{equation}
with $C \approx 6$, consistent with Refs.~\cite{ellenbroek2006critical, ellenbroek2009jammed}. We have checked that the fluctuations of $d_r(r)$ do not collapse as functions of $r \Delta Z$ (see Fig.~\ref{Fig:l_iso}(\textit{c,d})).

%\red{With both the screening length $\ell_{\rm s} = 1/\kappa_{\rm fit}$ (Fig.~2) and the isostatic length $\ell_{\rm iso}$ determined (Fig.~\ref{Fig:l_iso} and Eq.~\ref{eq:l_iso}), we can now compare them quantitatively. As expected in the main text and Ref.~\cite{jin2023intermediate}, the two lengths match  near the PE transition pressure $p_{\rm c}$, $\ell_{\rm s} = \ell_{\rm iso}$ (see Fig.~\ref{Fig:matching}). Note that $p_{\rm c}$ is determined independently in Fig.~2 and~3 using other criteria. Figure~\ref{Fig:matching} shows strong evidence that the PE transition results from  the combined effects of plastic screening and isostatic stability.}

%\begin{figure}[htbp]
%  \centering
%\includegraphics[width=0.5\linewidth]{fig11_match.pdf}
%    \caption{\red{{\bf Matching of the screening length $\ell_{\rm s}$ and the isostatic length $\ell_{\rm iso}$ at $p_{\rm c}$.} The data points of  $\ell_{\rm iso}$ are obtained from Fig.~\ref{Fig:l_iso}b, and the dashed line represents $\ell_{\rm iso} = 75 p^{-1/3}$. The data points of $\ell_{\rm s}$ are determined by $\ell_{\rm s} =  1/\kappa_{\rm fit}$ using $\kappa_{\rm fit}$ in Fig.~2a, and the horizontal line represents a constant $\ell_{\rm s}\approx 37$.
    %~\YF{$\approx 1/\kappa_1^*\approx \it{r}_{\rm out}/\pi$}.   The marked $p_{\rm c} \approx 7$ is estimated independently at the pressure where the power spectrum  is a power-law $S(f) \sim f^{-1.5}$ (see Fig.~3b).}     }   
%	\label{Fig:matching}
%\end{figure}

%%% Each figure should be on its own page
\clearpage

%S9
\section{Contact changes}
\label{SI: Contact changes}

\begin{figure}[htbp]
  \centering
\includegraphics[width=0.8\linewidth]{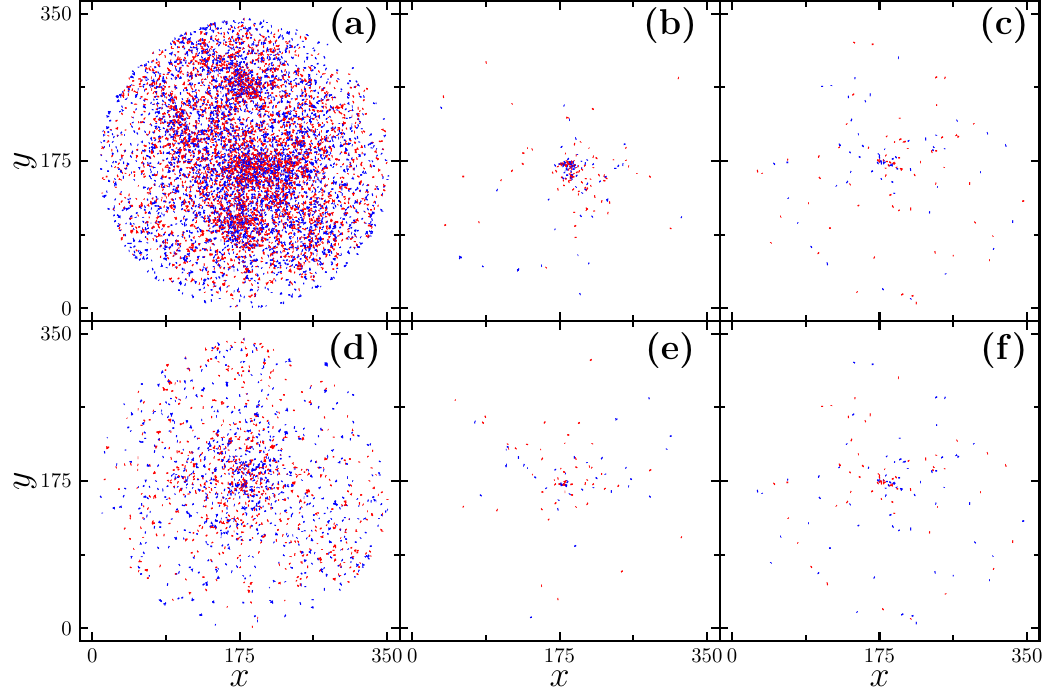}
    \caption{{\bf Contact changes due to (a-c) inflation and (d-f) deflation of the central disk.}
    The pressures are (a,d) $p=4.4 \times 10^{-2}$, (b,e) $p=5.2 \times 10^2$, (c,f) $p=8.6 \times 10^{2}$ (corresponding to the second to fourth rows in Fig.~1).  
    The inflation-deflation simulations are performed following the procedure $d_0 = 0 \to 0.1 \to 0$, and the configurations are equilibrated after the inflation and deflation. 
    The contact networks before and after the inflation/deflation  are compared.  
    The broken and created contacts are marked by blue and red points respectively.
    Note that there is no contact change for unjammed systems since $Z=0$.
    %Schematic representation of contact breaking (blue) and creation (red) following central disk inflation under different pressure.} (a)-(c) correspond to panels (d), (g), and (j) in Fig.~1 of the main text, while (d)-(f) correspond to the results when deflating the center particle back. Note that as the pressure increases, an increasing proportion of the contact changes revert to their initial states.
    }
	\label{Fig:contact_change}
\end{figure}

Figure~\ref{Fig:contact_change} shows that the inflation of the central particle causes multiple contact changes in our simulations. Thus the plastic-to-elastic (PE) transition does not correspond to the first contact change event analyzed in previous studies~\cite{van2014contact, van2016contact}.

\clearpage
% Produces the bibliography via BibTeX.

\end{document}